\documentclass[aps,prb,twocolumn,superscriptaddress]{revtex4}
\usepackage{graphicx}
\usepackage{amsmath}
\usepackage[colorlinks=true, citecolor=blue, urlcolor=blue, linkcolor=blue,bookmarks=false,hypertexnames=true]{hyperref} 
\usepackage{doi}

\newcommand{\ket}[1]{\left| #1 \right\rangle}

\begin{document}
\graphicspath{}

\title{Band nesting and  exciton spectrum in monolayer MoS$_{2}$}
\author{Maciej Bieniek}
\affiliation{Department of Physics, University of Ottawa, Ottawa, Ontario, Canada K1N 6N5}
\affiliation{Department of Theoretical Physics, Wroc\l aw University of Science and Technology, Wybrze\.ze Wyspia\'nskiego 27, 50-370 Wroc\l aw, Poland}

\author{Ludmi\l a Szulakowska}
\affiliation{Department of Physics, University of Ottawa, Ottawa, Ontario, Canada K1N 6N5}

\author{Pawe\l \ Hawrylak}
\affiliation{Department of Physics, University of Ottawa, Ottawa, Ontario, Canada K1N 6N5}

\date{\today}

\begin{abstract}
We discuss here the effect of band nesting and topology on the spectrum of excitons in a single layer of MoS$_2$, a prototype transition metal dichalcogenide material. We solve for the single particle states using the \itshape ab initio \upshape based tight-binding model containing metal $d$ and sulfur $p$ orbitals. The metal orbitals contribution evolving from $K$ to $\Gamma$  points results in conduction-valence band nesting and a set of second minima at $Q$ points in the conduction band. There are three $Q$ minima for each $K$ valley.  We accurately solve the Bethe-Salpeter equation  including  both $K$ and $Q$ points and obtain ground and excited exciton states. We determine the effects of the electron-hole single particle energies including band nesting, direct and exchange screened Coulomb electron-hole interactions and resulting topological magnetic moments on the exciton spectrum. The ability to control different contributions combined with accurate calculations of the ground and excited exciton states allows for the determination of the importance of different contributions and a comparison with effective mass and $k\cdot p$ massive Dirac fermion models.  
\end{abstract}

\maketitle

\section{Introduction}
There is currently great interest in van der Waals materials, including semiconductors, topological insulators, semimetals, superconductors and ferromagnets\cite{Splendiani_Galli_2010, Mak_Heinz_2010, Geim_Grigorieva_2013, Wang_Strano_2012, Chhowalla_Zhang_2013, Xu_Heinz_2014, Yu_Yao_2015, Novoselov_Neto_2016, Mak_Shan_2016, Schaibley_Xu_2016, Ren_Niu_2016, Wang_Urbaszek_2018, Jin_Heinz_2018, Mak_Shan_2018, Mueller_Malic_2018, Schneider_Urbaszek_2018, Scharf_Dery_2019}. The weak bonding of atomic layers in bulk materials allows to peel off single layers from the bulk and reassemble them into new combinations not found in nature\cite{Geim_Grigorieva_2013}. Here we focus on the understanding of a single atomic layer of MoS$_2$\cite{Splendiani_Galli_2010, Mak_Heinz_2010, Kadantsev_Hawrylak_2012}, a prototype of transition metal dichalcogenides (TMDs). Bulk MoS$_2$ is an indirect gap semiconductor, but when thinned down to a single layer becomes a direct gap material, with conduction band minima at six $K$ points. Hence, a single layer is an example of a true 2D semiconductor and one could hope to observe an ideal 2D exciton spectrum. Such ideal 2D spectrum would show an increase of exciton binding energy E$_\textrm{b}$  from E$_\textrm{b}$(3D)=1 Rydberg (Ry) to E$_\textrm{b}$(2D)=4 Ry and increase of excited state energy from -1/4 E$_\textrm{b}$(3D)  to -1/9 E$_\textrm{b}$(2D). A similar effort has been made in GaAs quantum wells, but the finite thickness of the quantum well and screening by the surrounding material prevented observation of an ideal 2D exciton spectrum\cite{Miller_Gossard_1981, Miller_Kleinman_1985}.  MoS$_2$ also differs from a generic 2D semiconductor in several ways.  As pointed out by Rytova\cite{Rytova_1967} and Keldysh\cite{Keldysh_1979}, due to 2D character of the semiconductor the screening of a 3D electron-hole attraction should be reduced, resulting in exciton with very large binding energy. The second difference between GaAs quantum well and MoS$_2$ layer is the presence of two nonequivalent valleys, with low energy spectra described by massive Dirac fermion (mDF) Hamiltonians. The topological nature of mDFs results in topological magnetic moments,  opposite in each valley. The massive Dirac fermion dispersion departs from the parabolic free electron or hole dispersion, the screening by 2D material differs from the bulk screening and the presence of topological moments results in complex exciton spectrum. The exciton in a massive Dirac fermion model is particularly interesting, but the model  misses an important ingredient of the MoS$_2$ bandstructure, the band nesting. As discussed by, e.g., Kadantsev et al. \cite{Kadantsev_Hawrylak_2012} there are 6 secondary minima in the conduction band at $Q$ points. The presence of $Q$ points in the conduction band is due to the mixing of different metal orbitals between conduction and valence bands and results in band nesting and strong light-matter interaction\cite{Bernardi_Grossman_2013}. Hence, each Dirac fermion at two nonequivalent points $K$ and $-K$ is surrounded by three $Q$ points, situation resembling the quark physics due to emerging SU(3) symmetry of those states. Hence, to understand the spectrum of the exciton in MoS$_2$ one needs to be able to control and turn on and off different contributions.  While there are several microscopic GW-BSE calculations of the exciton spectrum \cite{Ramasubramaniam_2012, Komsa_Krasheninnikov_2012, Sanchez_Wirtz_2013,  Qiu_Louie_2013, Huser_Thygesen_2013, Shi_Yakobson_2013, Bernardi_Grossman_2013, Conley_Bolotin_2013, Klots_Bolotin_2014, Soklaski_Yang_2014, Ugeda_Crommie_2014, Ye_Zhang_2014, Qiu_Loiue_2015, Latini_Thugesen_2015, Qiu_Louie_2016, Sanchez_Wirtz_2016, Echeverry_Gerber_2016, Robert_Urbaszek_2016, Despoja_Marusic_2016, Jornada_Louie_2017, Guo_Fleming_2019, Latini_Rubio_2019} we opt here for \itshape ab initio \upshape based tight-binding model of conduction and valence band states \cite{Bieniek_Hawrylak_2018}. This model allows us to understand and monitor contribution of different d-orbitals across the Brillouin zone, from $K$ to $Q$ to $\Gamma$ points. The contributions of different orbitals combined with accurate calculations of direct and exchange screened Coulomb matrix elements, and a highly converged solutions of Bethe-Salpeter equation, capture the $K$ and $Q$ valley contributions.  This approach allows us to investigate the role of different effects, from conduction band dispersion through $Q$ points, effect of different orbitals and topology on Coulomb matrix elements, to screening on the ground and excited exciton spectrum. Results of calculations are compared with a number of available experiments\cite{Chernikov_Heinz_2014, Goryca_Crooker_2019, Molas_Potemski_2019}.

From theoretical point of view, it is well-known that obtaining numerical solutions of Bethe-Salpeter equation is computationally challenging due to poor scaling with respect to increasing mesh of k-points discretizing the first Brillouin zone \cite{Qiu_Louie_2013, Jornada_Louie_2017}. This problem becomes even more severe when larger, experimentally relevant optical complexes \cite{Scrace_Hawrylak_2015, Jadczak_Hawrylak_2017, Jadczak_Bryja_2017, Jadczak_Hawrylak_2019} are considered in reciprocal space via generalized Bethe-Salpeter equations, e.g. for trions\cite{Druppel_Rohlfing_2017, Arora_Bratschitsch_2019, Zhumagulov_Perebeinos_2020} and biexcitons\cite{Steinhoff_Li_2018}. State of the art DFT+GW+BSE calculations allow to reach only 12x12 k-point grids \cite{Jornada_Louie_2017}. Density of those grids can be further increased under some approximations, usually related to interpolated\cite{Rohlfing_Louie_2000} and/or simplified Coulomb matrix elements and models of screening, reaching a record of 1024x1024 points using algorithms utilizing matrix product states and density matrix renormalization group framework\cite{Kuhn_Richter_2019, Kuhn_Richter_2019b}. 

The problems related to accurate calculation of the ground state of the exciton naturally affect calculations of exciton fine structure, which can be understood only approximately in terms of "Ising" excitons and spin splitting in both valance and conduction bands. For example, contradictory results of DFT studies of MoS$_2$ ground state have been reported \cite{Qiu_Loiue_2015, Echeverry_Gerber_2016, Malic_Berghauser_2018, Yu_Rubel_2019, Deilmann_Thygesen_2019}, some suggesting that MoS$_2$ has a ground exciton state that is indirect in exciton center-of-mass momentum\cite{Yu_Rubel_2019, Deilmann_Thygesen_2019}. In addition, excitonic spectrum depends heavily on screening of interactions due to dielectric environment \cite{Lin_Palacios_2014, Cho_Berkelbach_2018, Hsu_Shih_2019} and carrier-carrier screening for doped samples \cite{Hawrylak_1991, Chang_Reichman_2019}. 

It is also known that convergence of the calculations of the ground exciton state is challenging but achievable, however, accurate determination of energies and wavefunctions of excited exciton states is increasingly difficult with increasing exciton excited state energy. Theoretical models and calculations of exciton excited s-series, experimentally accessible in single photon emission/absorption experiments \cite{Chernikov_Heinz_2014, He_Mak_Shan_2014, Wang_Urbaszek_2015, Robert_Urbaszek_2018}, are usually based on model dispersion and screened Rytova-Keldysh interaction  \cite{Olsen_Thygessen_2016} or models using dielectric screening functions with various levels of sophistication \cite{Andersen_Thygesen_2015, Andersen_Thygesen_2015b, Molas_Potemski_2019}. State of art experiments, performed in magnetic fields allowing to identify signal from excited exciton states, were reported for different MX$_2$ compounds \cite{Stier_Crooker_2016, Mak_Shan_2018, Stier_Crooker_2018, Goryca_Crooker_2019, Delhomme_Faugeras_2019, Molas_Potemski_2019}. 

The dark excited exciton states in TMD's, e.g. 2p states in second excitonic shell \cite{Berkelbach_Reichman_2015}, also generated a lot of interest due to their novel topological properties\cite{Srivastava_Imamoglu_2015, Zhou_Xiao_2015, Wu_MacDonald_2015}. The predicted splitting of the p-shell could be understood in terms of  topological magnetic moments, consequence of  Berry's geometric curvature, acting on finite angular momentum states as an effective magnetic field \cite{Hichri_Goerbig_2019, Donck_Peeters_2019}. The same magnetic moments result in shift of the energy levels of s - series\cite{Trushin_Belzig_2018}, for which exciton's angular quantum number L is 0. The p-states can be probed in pump-probe experiments \cite{Pollmann_Huber_2015, Cha_Choi_2016, Steinleitner_Huber_2018, Berghauser_Malic_2018, Brem_Malic_2019, Merkl_Korn_2019, Yong_Wang_2019}, where 1s excitons are generated and transition from 1s to 2p states can be measured by a probe terahertz beam \cite{Berkelbach_Reichman_2015}.
 
In this work we construct a theory of exciton in MoS$_2$ starting with a 6-orbital \itshape ab initio \upshape based tight-binding model\cite{Bieniek_Hawrylak_2018}. We obtain the valence and conduction bands reproducing the \itshape ab initio \upshape results, including dispersion in the vicinity of K, Q and $\Gamma$ points. We fill all the single particle states in the first Brillouin zone of the valence band with electrons and construct electron-hole pair excitations on a grid of k-points in a single valley, $K$ and $-K$. We compute direct and exchange matrix elements describing electron-hole interaction on our numerical grid. We solve the Bethe-Salpeter equation with high accuracy obtaining exciton states for different approximations to energy dispersion and interactions. This  allows for discussion of different contributions to the n=1 to n=4 excitonic s-shells and the 2p shell in a computationally converged manner. 
    
We establish a connection between effective mass, massive Dirac fermion and tight-binding models of the electron-hole dispersion, and then using them rigorously on the same numerical grid defined for one valley, we study how they affect excitonic spectrum and renormalize energy levels toward "more than 2D" exciton. In this analysis we isolate the effect of three $Q$ points around the minimum of +K valley on the excitonic series. Next we show how exciton series gets renormalized back to more "3D-like" spectrum due to Rytova-Keldysh screening of electron-hole interaction. Next we discuss how form factors in Coulomb interaction, resulting from orbital structure of electron and hole Bloch wavefunctions, modifies the excitonic series. The comparison between form factors obtained from the massive Dirac fermion model and full microscopic tight-binding theory shows the limited validity of the Dirac fermion model  when used for the entire  valley of MoS$_2$. Next we confirm the topological splitting of the 2p states and determine how this splitting is affected by both screening of the Coulomb interactions and the presence of $Q$ points. We follow with discussion of the exciton fine structure due to spin-orbit coupling. Finally, we present a mechanism of bright to dark exciton ground state transition, a result of both the electron-hole exchange interaction and different effective masses of spin-split conduction bands of MoS$_2$. Finally, we find that the ground exciton state in MoS$_2$ is optically dark. 

Our paper is organized in the following way. In Section II we begin with discussion of the conduction and valence band electronic dispersion models, from our microscopic tight-binding Hamiltonian to the massive Dirac fermion theory and further to the effective mass model. In Section III A we re-derive Bethe-Salpeter equation for electron-hole pair excitations.  In Sections III B-E we construct Coulomb matrix elements from  Bloch wavefunctions with realistic atomic-like orbitals to obtain microscopic form factors of both electron-hole attractive direct and repulsive exchange interactions. In Section III F we analyze the spectrum of ideal 2D exciton and compare analytical and numerical results. In Section III G we deal with Coulomb singularities in electron-hole interactions. Section IV A  describes  the single valley exciton and  Section IV B exciton fine structure due to  spin and valley. In Section V we discuss the exciton spectrum, starting with the effect of Q - points in Section V A, screening of Coulomb interactions  in the Rytova-Keldysh approximation in Section V B and renormalization of the 2D exciton spectrum towards "3D-like" exciton is then discussed in Section V C. The role of topology in form factors of electron-hole interactions, spin-orbit splitting of conduction bands and exchange interaction on exciton fine structure are then discussed in Sections V D-F. We conclude in Section VI with brief summary. We note that reader less interested in technical aspects of tight-binding and Bethe-Salpeter methodology used here can skip Sections II-IV and start from Section V, where description of numerical results begins.

\section{Single particle spectrum in MoS$_2$}

\begin{figure}[ht]
\centering
\includegraphics[width=0.5\textwidth]{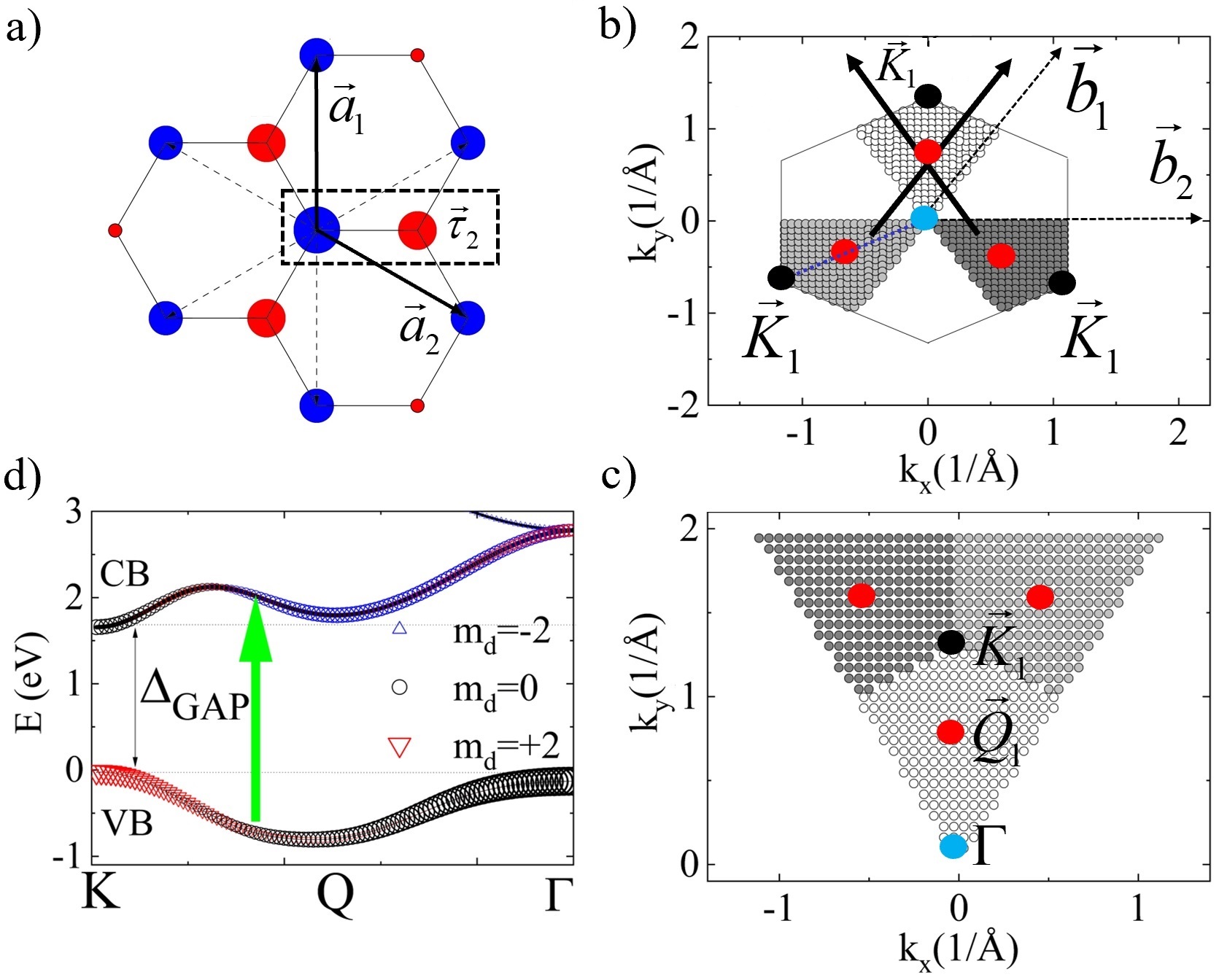}
\caption{(a) MoS$_2$ structure with unit cell shown inside dashed line. Blue and red points represent metal Mo and sulfur dimer S$_2$ positions, respectively. (b) Hexagonal Brillouin zone showing equivalent K-points and choice of points corresponding to one valley.  (c) Construction of triangle around K-point representing single valley. (d) Valence and conduction bands dispersion along K-Q-$\Gamma$ line, shown as dotted line on (b). Green arrow represents band nesting transition energy. Size of symbols denotes orbital contribution from Mo $m_d=\pm 2, 0$ orbitals to bands at given k points.} 
\label{fig1}
\end{figure}

We describe here the valence and conduction bands of MoS$_2$. We begin with the description of atomic  lattice of monolayer  2H - MoS$_2$, defined by vectors $a_1= d_{\parallel}\left(0,\sqrt{3}\right)$, $a_2= d_{\parallel}/2(3,-\sqrt{3})$, with top view  shown in Figure \ref{fig1}(a). Blue and red dots represent positions of molybdenum Mo atoms and sulfur dimers S$_2$, respectively. The distance between Mo atom and S$_2$ dimer center in the z=0 plane, $d_{\parallel}$, is set to 1.84 \AA. Here dashed line denotes the choice of a unit cell and vector $\tau_2=\left(d_{\parallel},0\right)$ denotes position of a sulfur dimer center inside  the unit cell ($\tau_1=0$). Using the usual definition $\exp(i\vec{a}_i\cdot \vec{b}_j)=\delta_{ij}$, reciprocal lattice vectors $\vec{b}_1$ and $\vec{b}_2$ defining hexagonal Brillouin zone (BZ) are constructed, as presented in Fig. \ref{fig1}(b). In Fig. \ref{fig1}(b) the three equivalent K$_1$ points are located at the vertices of the hexagon. It is important to note how we select  a subset of all 
k-space points inside the hexagon  associated with +K valley. These points create three "kites" around the $\Gamma$ point and are related by C$_3$ symmetry. Arrows on Fig. \ref{fig1}(b) show how these three kite regions can be translated by reciprocal lattice vectors to the neighborhood of one of the K$_1$ points, creating a triangle with $+K $ point in the center, as shown in Fig. \ref{fig1}(c). This triangle contains all points in k-space associated with valley $+K$. In the vicinity of $+K$ minimum we have circles describing massive Dirac fermions with constant energy. A second triangle around nonequivalent $-K$ valley can be created in analogous way. 

We now move to construct a tight-binding (TB) theory of the electronic structure, as discussed in detail in Ref. \onlinecite{Bieniek_Hawrylak_2018}. We write electron wavefunction as a linear combination of atomic orbitals 
$\varphi_{\alpha\mu}$ as follows: 
\begin{equation}
\begin{split}
&\Psi_{n}\left(\vec{k},\vec{r}\right)=e^{i\vec{k}\cdot\vec{r}}u_{n}\left(\vec{k},\vec{r}\right)= \\
&e^{i\vec{k}\cdot\vec{r}}\frac{1}{\sqrt{N_{uc}}}\sum_{i=1}^{N_{uc}}\sum_{\alpha=1}^{2}\sum_{\mu=1}^{3}e^{-i\vec{k}\cdot\left(\vec{r}-\vec{U}_{i}-\vec{\tau}_{\alpha}\right)}v_{\alpha\mu}^{(n)}\varphi_{\alpha\mu}\left(\vec{r}-\vec{U}_{i}-\vec{\tau}_{\alpha}\right),
\end{split}
\label{eq1}
\end{equation}
where $\it{n}$ denotes band and $u_n$ are periodic Bloch functions. Coefficients $\nu_{\alpha\mu}$ are k - dependent functions obtained from the TB Hamiltonian for atom type $\alpha$ and orbital $\mu$ $(\textrm{Mo} \rightarrow \alpha=1,\mu \leftrightarrow m_d=\pm 2,0$; $\textrm{S}_2 \rightarrow \alpha =2, \mu \leftrightarrow m_p=\pm 1,0)$. The atomic functions $\varphi_{\alpha\mu}$ are localized in unit cells centered around  $U_i$,  unit cell coordinates,  with $\tau_\alpha$ being atomic  positions inside each unit cell. These orbitals are modelled using Slater-type, single-$\zeta$ basis with parameters from Refs. \onlinecite{Clementi_Raimondi_1963, Clementi_Reinhardt_1967}. 

Using the atomic orbitals, the minimal tight-binding Hamiltonian in block form emphasizing Mo-S$_2$ orbital interaction can be written as: 
\begin{gather}
\hat{H}^{TB}\left(\vec{k} \right)=
\begin{bmatrix}
H_{Mo-Mo} & H_{Mo-S_2} \\
H_{Mo-S_2}^{\dagger} & H_{S_{2}-S_{2}}  \\ 
   \end{bmatrix},
\label{6x6}
\end{gather}
\begin{gather*}
H_{Mo-Mo}=
\begin{bmatrix}
^{E_{m_{_{d}}=-2}}_{{+}W_{1}g_{0}(\vec{k})}& W_{3}g_{2}(\vec{k}) & W_{4}g_{4}(\vec{k}) \\
& ^{E_{m_{_{d}}=0}}_{{+}W_{2}g_{0}(\vec{k})} & W_{3}g_{2}(\vec{k})\\ 
&  & ^{E_{m_{_{d}}=2}}_{{+}W_{1}g_{0}(\vec{k})} 
   \end{bmatrix},
\end{gather*}
\begin{gather*}
H_{S_{2}-S_{2}}=
\begin{bmatrix}
^{E_{m_{_{p}}=-1}}_{{+}W_{5}g_{0}(\vec{k})} & 0 & W_{7}g_{2}(\vec{k}) \\
& ^{E_{m_{_{p}}=0}}_{{+}W_{6}g_{0}(\vec{k})} & 0 \\
& & ^{E_{m_{_{p}}=1}}_{{+}W_{5}g_{0}(\vec{k})} \\
   \end{bmatrix},
\end{gather*}
\begin{gather*}
H_{Mo-S_{2}}=
\begin{bmatrix}
V_{1}f_{-1}(\vec{k}) & -V_{2}f_{0}(\vec{k}) & V_{3}f_{1}(\vec{k}) \\
-V_{4}f_{0}(\vec{k}) & -V_{5}f_{1}(\vec{k}) & -V_{4}f_{-1}(\vec{k}) \\
-V_{3}f_{1}(\vec{k}) & -V_{2}f_{-1}(\vec{k}) & V_{1}f_{0}(\vec{k}) 
   \end{bmatrix},
\end{gather*}
where $f$ and $g$ are k-dependent functions multiplied by amplitudes $V$, $W$. These amplitudes parametrize nearest and next-nearest neighbor interactions, respectively (see Appendix A and B in Ref. \onlinecite{Bieniek_Hawrylak_2018}).  We note that $\zeta$ parameters entering atomic orbitals $\varphi$ are independent from Slater-Koster integrals parametrizing TB model, as usually assumed \cite{Slater_Koster_1954}. Under the approximation of orthogonal basis of atomic orbitals, the eigenproblem for TB energies and wavefunction coefficients is given by
\begin{equation}
\hat{H}^{TB}\left(\vec{k}\right)\bar{v}^{(n)}\left(\vec{k}\right)=\varepsilon^{TB}_{n}\left(\vec{k}\right)\bar{v}^{(n)}\left(\vec{k}\right).
\label{eq3}
\end{equation}
The parameters of the TB Hamiltonian are obtained from the \itshape ab initio \upshape calculations. The energy levels and wavefunctions are obtained by diagonalizing the TB Hamiltonian on a lattice of k-points. Figure \ref{fig1}(d) shows the dispersion of the conduction $E_{CB}(k)$ and valence band $E_{VB}(k)$ energy levels obtained in our \itshape ab initio \upshape based TB model. In the conduction band we see a  minimum at $K$ and a secondary minimum at $Q$ points. The energy minimum at $Q$ point implies that the conduction and valence bands run parallel in energy as a function of the wavevector $k$, resulting in conduction-valence band nesting and enhanced joint optical density of states. 

If we are interested in the vicinity of $+K$ point, the results of the TB model can be well approximated by a massive Dirac fermion (mDF) Hamiltonian,
\begin{gather}
\hat{H}^{mDF}\left(\vec{q}\right)=
\begin{bmatrix}
\Delta/2 & \hbar v_{\textrm{F}}\left(iq_{x}-q_{y}\right)  \\
\hbar v_{\textrm{F}}\left(-iq_{x}+q_{y}\right) & -\Delta/2 
\end{bmatrix},
\label{eq4}
\end{gather}
where $\vec{q}$ vectors are measured from $K$ point ($\vec{k}=\vec{K}+\vec{q}$) and $\hbar v_{\textrm{F}}=at=\left(3d_{||}/2\right)t=3.51\textrm{ eV\AA}$. For this model, an analytical formula for conduction (+) and valence (-) energy dispersion of massive Dirac fermions can be obtained as $\varepsilon^{mDF}_{\pm}=\pm\sqrt{\left(\Delta/2\right)^{2}+(atq)^{2}}$. Furthermore, in the limit of large energy gap, $(2atq/\Delta)^{2}<<1$, the dispersion of quasiparticles in the  mDF model can be further simplified to parabolic dispersion,
\begin{equation}
\varepsilon^{mDF}_{\pm}=\pm\frac{\Delta}{2}\sqrt{1+\frac{4a^{2}t^{2}}{\Delta^{2}}q^{2}} \approx \pm\left(\frac{\Delta}{2}+\frac{\hbar^{2}q^{2}}{2m^{*}}\right)     ,
\label{eq5}
\end{equation}
with effective mass given by $m^{*}=\hbar^{2}\Delta/(2a^{2}t^{2})$. We note that while the two models offer significant simplicity, they  miss a crucial effect of band nesting around Q point. This is shown in Fig. \ref{fig2}(a) which shows  the dispersion of  electron - hole complexes  $\Delta E \left(k \right)=\varepsilon_{\textrm{CB}}(k)-\varepsilon_{\textrm{VB}}(k) $  for a parabolic model, massive DF model and full microscopic TB model.

\section{Bethe-Salpeter equation for valley excitons}

\subsection{Bethe-Salpeter equation}

\begin{figure}[ht]
\centering
\includegraphics[width=0.5\textwidth]{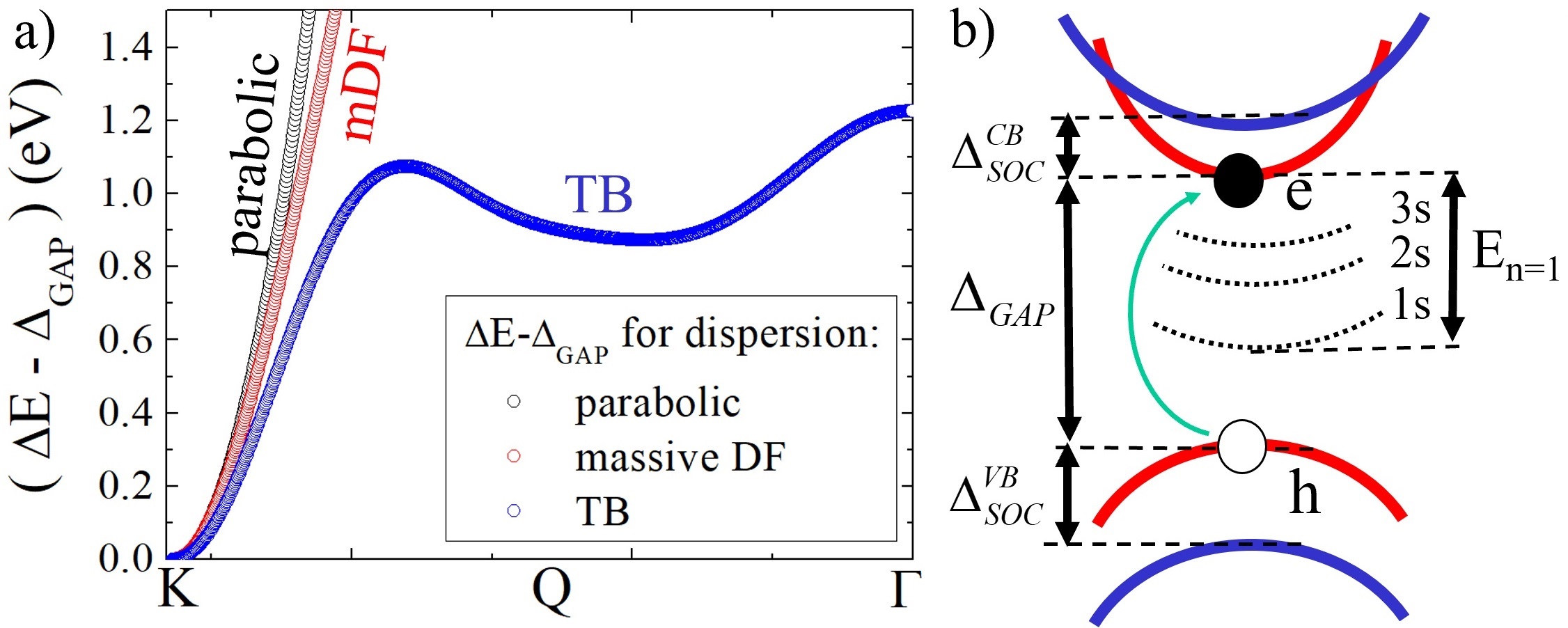}
\caption{(a) Three different models of single-particle electron-hole transition energies $\Delta E(k)$ along K-Q-$\Gamma$, showing pronounced density of states around Q point for tight-binding (TB) model absent in parabolic band and massive Dirac fermion models. (b) Schematic representation of spin-dependent bands close to the K-point with example of vertical electron-hole excitation. Correlated states of these excitations correspond to exciton series with binding energy $E_n$. $\Delta_{\textrm{GAP}}$ and $\Delta_{\textrm{SOC}}^{\textrm{CB/VB}}$ denote fundamental band-gap and spin splitting in valence and conduction band at K-point. } 
\label{fig2}
\end{figure}

After defining single-particle  states in the first Brillouin zone, we fill up all the states in the valence band and form a single Slater determinant as the ground state of non-interacting Kohn-Sham particles $\ket{\textrm{GS}}$. We next turn on the remaining electron-electron interactions. We form the valley exciton state $\ket{X,Q_{\textrm{CM}}}_{n} $ as a linear combination of electron-hole excitations, $c^{\dagger}_{c,k+Q_{\textrm{CM}},\sigma}c_{v,k,\sigma}\ket{\textrm{GS}}$, out of the ground state, where $c^{\dagger}_{c,k+Q_{\textrm{CM}},\sigma}$ creates an electron in conduction band state $\ket{c,k+Q_{\textrm{CM}},\sigma}$ and $c_{v,k,\sigma}$ annihilates an electron in the valence band state $\ket{v,k,\sigma}$:
\begin{equation}
\ket{X,Q_{\textrm{CM}}}_{n}=\sum_{k}^{1^{st}BZ} A^{Q_{\textrm{CM}}}_{n}\left(\vec{k}\right) c^{\dagger}_{c,k+Q_{\textrm{CM}},\sigma} c_{v,k,\sigma}\ket{\textrm{GS}}. 
\label{eq6}
\end{equation}
Here $A^{Q_{\textrm{CM}}}_{n}$ are electron-hole complex amplitudes in the exciton wavefunction $\ket{X,Q_{\textrm{CM}}}_{n}$ in exciton state 'n' with total, center-of-mass, momentum $Q_{\textrm{CM}}$. For optically relevant excitons $Q_{\textrm{CM}}=0$ (we drop index $Q_{\textrm{CM}}$ from now on). The electron-hole excitations are not the eigenstates of the interacting Hamiltonian and are mixed by electron-electron interactions. We write the interacting Hamiltonian for states $\ket{i}=\ket{b,k,\sigma}$, $b=v,c$, as
\begin{equation}
\hat{H}=\sum_{i}\varepsilon_{i}c_{i}^{\dagger}c_{i}+\frac{1}{2}\sum_{ijkl}\left\langle i\left| j \left|V \right| k \right| l \right\rangle c_{i}^{\dagger}c_{j}^{\dagger}c_{k}c_{l}
\label{eq7}
\end{equation}
where $\left\langle i\left| j \left|V \right| k \right| l \right\rangle$ are electron-electron interaction matrix elements measured from the mean field. Exciton states are obtained by solving the exciton equation $\hat{H}\ket{X}_{n}=E_{n}\ket{X}_{n}$, where $E_{n}$ are exciton energies and $\ket{X}_{n}$ are exciton states (see schematic on Fig. \ref{fig2}(b)). The resulting Bethe-Salpeter equation for exciton amplitudes and energies is given by
\begin{equation}
\begin{split}
&\left[\Delta E \left(\vec{k} \right) - \Delta_{\textrm{GAP}} + \Sigma\left(\vec{k} \right) \right]A_{n}\left(\vec{k} \right) + \\
&\sum_{\vec{k}'}\begin{bmatrix}
-\left\langle v,\vec{k}'| c,\vec{k} | V | c,\vec{k'} | v,\vec{k} \right\rangle  \\
+\left\langle v,\vec{k}'| c,\vec{k} | V | v,\vec{k} | c,\vec{k}' \right\rangle  \\
\end{bmatrix}A_{n}\left(\vec{k}'\right)=E_{n}A_{n}\left(\vec{k}\right),
\end{split}
\label{eq8}
\end{equation}
where $\Delta E \left(\vec{k} \right)=\varepsilon_{\textrm{CB}}(\vec{k})-\varepsilon_{\textrm{VB}}(\vec{k}) $ is the energy difference between uncorrelated electron - hole pairs at given wavevector $\vec{k}$. Summation over bands is simplified to one valence and one conduction band, as assumed by form of Eq. (\ref{eq6}). For clarity, at first we neglect the spin-orbit interaction induced spin splitting of bands and we  take spinless electron wavefunctions due to their weak, Zeeman-like, dependence on spin-orbit coupling. For this study of excitons the electron and hole self-energies $\Sigma \left(\vec{k} \right)$ are treated as a k-independent quantities and included in the renormalized energy gap. The summation over $\vec{k}'$ in Eq. (\ref{eq8}) is understood as over all reciprocal lattice points, with number equal to number of atoms in a crystal. To make problem numerically tractable, usual transition from sum to integral, $\sum _{k'}\rightarrow \frac{S}{\left(2\pi\right)^{2}}\iint_{BZ}d^{2}k'$ , is performed, where S is the crystal area. 
In Eq. (\ref{eq8}) we find two forms of interactions between electron in the conduction band and a missing electron, a hole, in the valence band, a direct attractive interaction $-\left\langle v,\vec{k}'\left| c,\vec{k} \right|V \left| c,\vec{k'} \right| v,\vec{k} \right\rangle$ and repulsive, exchange $+\left\langle v,\vec{k}'\left| c,\vec{k} \right|V \left| v,\vec{k} \right| c,\vec{k}' \right\rangle$ interaction. Note that careful evaluation of Coulomb matrix elements leads to direct and exchange interaction  only  in band indices, but not in momentum $k$ indices.

\subsection{Direct Coulomb matrix elements}

The evaluation of direct and exchange Coulomb matrix elements \cite{Sundararaman_Arias_2013, Rosner_Wehling_2015} is critical to the results presented here and hence we provide detailed analysis in what follows. Direct Coulomb matrix elements are evaluated in the basis of only electrons, not holes. In this language, scattering of electron in CB at k (c,k) and valence hole at k (v,k) into electron in CB at k' (c,k') and hole in the valence band at k' (h,k') is equivalent to scattering of two electrons at (c,k')(v,k) to states of two electrons (c,k)(v,k') in Eq. (\ref{eq8}). The matrix elements are hence constructed from electron wavefunctions as
\begin{equation}
\begin{split}
&\left\langle v,\vec{k}'\left| c,\vec{k} \right|V \left| c,\vec{k'} \right| v,\vec{k} \right\rangle= \\
& \iint_{R^{3}} d^{3}rd^{3}r' V^{\textrm{3D}}\left(\vec{r}^{\textrm{ 3D}}-\vec{r}^{\textrm{ 3D}'} \right) \times \\
&\Psi^{*}_{v}\left(\vec{k'},\vec{r},z \right) \Psi^{*}_{c}\left(\vec{k},\vec{r'},z' \right)\Psi_{c}\left(\vec{k'},\vec{r'},z' \right) \Psi_{v}\left(\vec{k},\vec{r},z \right),
\end{split}
\label{eq9}
\end{equation}
where $V^{\textrm{3D}}$ is the effective electron-electron interaction. In Eq. (\ref{eq9})  we explicitly separated two dimensional vector $\vec{r}$ (in which crystal is periodic) and out of plane coordinate $z$, with $d^{3}r=d^{2}rdz$. Substituting the Bloch form of the wavefunctions and re-grouping them we get Eq. (\ref{eq9}) equal to 
\begin{equation}
\begin{split}
&\iint_{R^{3}} d^{3}rd^{3}r' e^{i\left(\vec{k}-\vec{k'}\right)\cdot\left(\vec{r}-\vec{r'} \right)} V^{\textrm{3D}}\left(\vec{r}^{\textrm{ 3D}}-\vec{r}^{\textrm{ 3D}'} \right) \times \\
&u^{*}_{v}\left(\vec{k}',\vec{r},z \right)u_{v}\left(\vec{k},\vec{r},z \right) \times u^{*}_{c}\left(\vec{k},\vec{r'},z' \right)u_{c}\left(\vec{k'},\vec{r'},z' \right).
\end{split}
\label{eq10}
\end{equation}
Next, we use 2D Fourier transform $V^{\textrm{3D}}(q)$ of 3D real space Coulomb interaction
\begin{equation}
\begin{split}
& V^{\textrm{3D}}\left(\vec{r}^{\textrm{ 3D}}-\vec{r}^{\textrm{ 3D}'} \right)=\\
&\frac{1}{\left(2\pi\right)^{2}}\iint_{-\infty}^{\infty}d^{2}q V^{\textrm{2D}}(q)e^{-|z-z'|\cdot \left|\vec{q}\right|} e^{i\vec{q}\cdot \left(\vec{r}-\vec{r'}\right)}
\end{split}
\label{eq11}
\end{equation}
and write products of Bloch wavefunctions as 2D Fourier series
\begin{equation}
u^{*}_{v}\left(\vec{k}',\vec{r},z \right)u_{v}\left(\vec{k},\vec{r},z \right) \equiv \rho_{vv}^{\vec{k'}\vec{k}}\left(\vec{r},z\right)=\sum_{\vec{G}}e^{i\vec{G}\cdot\vec{r}}\tilde{\rho}_{vv}^{\vec{k'}\vec{k}}\left(\vec{G},z\right).
\label{eq12}
\end{equation}
Fourier coefficients of pairs of Bloch functions, i.e. of pair densities, are given by
\begin{equation}
\tilde{\rho}_{vv}^{\vec{k'}\vec{k}}\left(\vec{G},z\right)=\frac{1}{S}\iint_{R^{2}} d^{2}r e^{-i\vec{G}\cdot\vec{r}}u^{*}_{v}\left(\vec{k}',\vec{r},z \right)u_{v}\left(\vec{k},\vec{r},z \right),
\label{eq13}
\end{equation}
where $S$ is crystal area and integration is performed over whole 2D plane $R^{2}$. Substituting Eq. (\ref{eq11}) and (\ref{eq12}) in Eq. (\ref{eq10}), integrating out delta functions, changing continuous delta to discrete one $\delta\left(\vec{G'}+\vec{G}\right)\rightarrow \frac{S}{\left(2\pi\right)^{2}}\delta_{\vec{G'}+\vec{G}}$ and using 2D Fourier transform of bare Coulomb interaction $V^{\textrm{2D}}(q)=\frac{e^{2}}{4\pi\varepsilon_{0}}\frac{2\pi}{q}$, we obtain final expression for direct matrix element (with coefficient $S/(2\pi)^{2}$ resulting from sum to integral transition)
\begin{equation}
\begin{split}
&V^{D}\left(\vec{k},\vec{k'}\right) = \frac{S}{\left(2\pi\right)^{2}}\left\langle v,\vec{k}'\left| c,\vec{k} \left|V \right| c,\vec{k'} \right| v,\vec{k} \right\rangle=\\
&\gamma\sum_{\vec{G}}\frac{F^{D}\left(\vec{k},\vec{k'},\vec{G}\right)}{\left|\vec{k'}-\vec{k}-\vec{G} \right|},
\end{split}
\label{eq14}
\end{equation}
where $\gamma=e^{2}/\left(8\pi^{2}\varepsilon_{0}\right)$ and direct Coulomb interaction form factor $F^{D}$ is given by:
\begin{equation}
\begin{split}
&F^{D}\left(\vec{k},\vec{k'},\vec{G}\right)=\\
&\int dz \int dz' \rho_{vv}^{\vec{k'}\vec{k}}\left(\vec{G},z\right)\rho_{cc}^{\vec{k}\vec{k'}}\left(-\vec{G},z'\right)e^{-\left|z-z'\right|\cdot\left|\vec{k'}-\vec{k}-\vec{G}\right|}.
\end{split}
\label{eq15}
\end{equation}
Pair densities (also called co-densities) can be evaluated by using explicit form of the Bloch wavefunctions, e.g., as
\begin{equation}
\begin{split}
&\rho_{vv}^{\vec{k'}\vec{k}}\left(\vec{G},z\right)=\frac{1}{N_{\textrm{UC}}}\sum_{\alpha,\beta=1}^{2}\sum_{\mu,\nu=1}^{3}\left[v^{\textrm{VB}}_{\alpha\mu}\left(\vec{k'}\right)\right]^{*}v^{\textrm{VB}}_{\beta\nu}\left(\vec{k}\right)\times\\
& \sum_{i,j=1}^{N_{\textrm{UC}}}\exp\left[-i\vec{k'}\cdot\left(\vec{U}_{i}+\vec{\tau}_{\alpha}\right)+i\vec{k}\cdot\left(\vec{U}_{j}+\vec{\tau}_{\beta}\right)\right]\times \\
&\iint_{R^{2}} d^{2}r \bigg\{ \exp\left[-i\left(\vec{G}-\vec{k'}+\vec{k}\right)\cdot\vec{r}\right]\times\\
&\varphi_{\alpha\mu}\left(\vec{r}-\vec{U}_{i}-\vec{\tau}_{\alpha},z\right)^{*}\varphi_{\beta\nu}\left(\vec{r}-\vec{U}_{j}-\vec{\tau}_{\beta},z\right) \bigg\}.
\end{split}
\label{eq16}
\end{equation}
The number of unit cells $N_{\textrm{UC}}$  in Eq. (\ref{eq16}) plays the role of convergence parameter and we use $N_{\textrm{UC}}=7$ (central unit cell + 6 nearest neighbor unit cells) for converged results. Examples of z-dependence of pair densities are given in Appendix D. Two dimensional in-plane integrals and $z$, $z'$ integrations are carried out numerically. Finite summation over $\vec{G}$ vectors is also performed up to $G_{\textrm{cutoff}}$, as discussed in Appendix E.

\subsection{Choice of Gauge in Coulomb matrix elements}
We note that in general, direct electron - hole Coulomb matrix elements are complex, gauge - dependent quantities. Even though B.-S. eigenproblem is gauge independent, the choice of phase for coefficients $v$ in Eq. (\ref{eq16}) affects the symmetry of exciton states, as found in Ref. \onlinecite{Zhou_Xiao_2015}. To obtain ground state excitons with 1s symmetry, we follow the gauge introduced by Rohlfing and Louie\cite{Rohlfing_Louie_2000}, for which a sum of imaginary parts of coefficients $v$ is set to zero, $\sum_{\alpha=1}^{2}\sum_{\mu=1}^{3}\textrm{Im}v_{\alpha\mu}^{(n)}=0$. The global phase is chosen such that $\textrm{Im }v_{12}=0$, i.e. , the coefficient for Mo $m_{d}=0$ orbital is set to be real.

\subsection{Exchange Coulomb matrix elements}

Returning to Coulomb matrix elements, expressions for exchange matrix elements are given by
\begin{equation}
\begin{split}
&V^{X}\left(\vec{k},\vec{k'}\right) = \frac{S}{\left(2\pi\right)^{2}}\left\langle v,\vec{k}'\left| c,\vec{k} \left|V \right| v,\vec{k} \right| c,\vec{k'} \right\rangle=\\
& \gamma \sum_{\vec{G}\neq 0} \frac{F^{X}\left(\vec{k},\vec{k'},\vec{G}\right)}{\left|\vec{G} \right|},
\end{split}
\label{eq17}
\end{equation}
with form factors given by
\begin{equation}
\begin{split}
&F^{X}\left(\vec{k},\vec{k'},\vec{G}\right)=\\
&\int dz \int dz' \rho_{vc}^{\vec{k'}\vec{k'}}\left(\vec{G},z\right)\rho_{cv}^{\vec{k}\vec{k}}\left(-\vec{G},z'\right)e^{-\left|z-z' \right|\cdot\left| \vec{G} \right|}.
\end{split}
\label{eq18}
\end{equation}
At this point we note   the $G=0$ singularity in $V^{X}$ in Eq. (\ref{eq17}). However, taking a limit  
\begin{equation}
\lim_{\vec{G}\to 0}\frac{\int dz \int dz' \rho_{vc}^{\vec{k'}\vec{k'}}\left(\vec{G},z\right)\rho_{cv}^{\vec{k}\vec{k}}\left(-\vec{G},z'\right)}{\left| \vec{G} \right|}=0,
\label{eq19}
\end{equation}
we find that this G=0 singular term does not contribute to $V^X$ and can be excluded from summation over G vectors in Eq. (\ref{eq17}).

\subsection{Effect of screening of Coulomb interactions}
Formulas for $V^{D}$ and $V^{X}$ are for bare, unscreened direct and exchange 3D Coulomb interactions. The electrons in 2D materials and in the substrate screen Coulomb interactions. In the following work we study two models of screening.
In the first "static" approximation the MoS$_{2}$ is embedded in a bulk material with dielectric constant $\varepsilon_{r}^{stat.}$ and the 3D Coulomb interaction $V(r,r')=e^2/(4\pi\varepsilon_{0}|r-r'|)$ is screened by a  dielectric constant $\varepsilon_{r}^{stat.}$, i.e.  $V(r,r')=e^2/(\varepsilon_{r}^{stat.} \cdot 4\pi\varepsilon_{0}|r-r'|$). Alternatively, Coulomb matrix elements are simply divided by the dielectric constant $\varepsilon_{r}^{\textrm{stat.}}$ as $V^{D/X}_{stat.}(q)=V^{D/X}_{bare}(q) / \varepsilon_{r}^{\textrm{stat.}}$. The second approximation is the Rytova-Keldysh screening\cite{Rytova_1967, Keldysh_1979, Cudazzo_Rubio_2011}, in which case we use $V^{D/X}_{R-K}(q)=V^{D/X}_{bare}(q) / \left[\varepsilon_{r}^{\textrm{R-K}}\left(1+2\pi \alpha \left|\vec{q}\right| \right) \right]$ , where $\alpha$ is the 2D material electron polarizability,  treated as a parameter here. $\varepsilon_{r}^{\textrm{R-K}}=\left(\varepsilon_{1}+\varepsilon_{3} \right)/2$ , in contrast with $\varepsilon_{r}^{\textrm{stat.}}$, describes effectively dielectric properties of surrounding materials / vacuum. We study the case of uncapped MoS$_{2}$ on SiO$_{2}$\cite{Tamulewicz_Gotszalk_2019, Jadczak_Bryja_2017}, therefore  $\varepsilon_{3}=1$ and $\varepsilon_{3}=4$ parameters are taken. We note that more advanced models of screening  do not affect significantly  exciton spectra as discussed in, e.g. Ref. \onlinecite{Berkelbach_Reichman_2018}.

In Appendix F we describe a simplified approach to direct electron-hole matrix elements, in which essentially we simplify interaction form factor to 1 and choose from $\vec{G}$ summation one $G$ that is minimizing $|\vec{k'}-\vec{k}-\vec{G}|$ vector, giving us interaction equivalent to the one usually assumed in massive Dirac fermion / parabolic dispersion theories around $K$ point with the form $1/|\vec{q}-\vec{q'}|$, where $q$ is a distance from $K$ point in k-space.

\subsection{Ideal 2D exciton}

We would like to compare excitons in MoS$_2$ with ideal 2D excitons and use the 2D exciton spectrum as a test of numerical accuracy.  Lets consider the ideal 2D exciton problem, electrons and holes with parabolic dispersion interacting via statically screened Coulomb interaction without electron-hole exchange interaction and neglecting self-energy $\Sigma\left(\vec{k}\right)$.  In this case the Fourier transform of Bethe-Salpeter equation, Eq.(\ref{eq8}), reduces to the well-known 2D hydrogen problem for electron-hole pair 
\begin{equation}
\hat{H}=\Delta_{\textrm{GAP}}-\frac{\hbar^{2}\nabla^{2}_{e}}{2m_{e}^{*}}-\frac{\hbar^{2}\nabla^{2}_{h}}{2m_{h}^{*}}-\frac{e^{2}}{4\pi\varepsilon_{0}\varepsilon_{r}^{stat.}\left|\vec{r_{e}}-\vec{r_{h}} \right| },
\label{eq20}
\end{equation}
which, in excitonic effective Rydberg and Bohr radius units $M=m_{e}^{*}+m_{h}^{*}$, $\mu=\left(1/m_{e}^{*} + 1/m_{h}^{*} \right)^{-1}$, $\varepsilon=4\pi\varepsilon_{0}\varepsilon_{r}^{stat.}$, $\hbar=2\mu=e^{2}/\left(2\varepsilon\right)=1$, Ry$^{\mu}=\mu e^{4}/\left(2\hbar^{2} \varepsilon^{2}\right)$, $a_{0}^{\mu}=\varepsilon \hbar^{2}/\left(\mu e^{2} \right)$, center-of-mass $\vec{R}=\left(m_{e}^{*}\vec{r_{e}}+m_{h}^{*}\vec{r_{h}}\right)/\left(m_{e}^{*}+m_{h}^{*}\right)$ and relative motion coordinates $\vec{r}=\vec{r_{e}} - \vec{r_{h}}$, can be written as
\begin{equation}
\hat{H}=\Delta_{\textrm{GAP}}-\frac{1}{M/\mu}\nabla^{2}_{R}-\nabla_{r}^{2}-\frac{2}{|r|}.
\label{eq21}
\end{equation}
Eq. (\ref{eq21}) can be solved using transformation to parabolic coordinates \cite{Hawrylak_Grabowski_94}, which maps the Coulomb problem for infinitely many bound states to a spectrum of 2D harmonic oscillator. In Ref. \onlinecite{Hawrylak_Grabowski_94} the solution is written in terms of 2D harmonic oscillator coordinates $n,m$ and yields $E_{nm}=-4/(n+m+1)^{2}$ Ry$^{\mu}$ with $n-m=\pm 2p, p=0,1,2$. This is equivalent to the following series of states:
\begin{equation}
E_{n}=-\frac{1}{\left(n-\frac{1}{2}\right)^{2}}\left[\textrm{Ry}^\mu\right], \qquad n=1,2, ..., \infty
\label{eq22}
\end{equation}
with degeneracies being same as every second shell of 2D harmonic oscillator. Hence, lowest energy state is nondegenerate with energy $E_{n=1}=-4$. The second state has three-fold degeneracy (2s, 2p$_x$, 2p$_y$) and energy $E_{n=2}=-4/9$, etc. We see that the excited state has energy equal only 1/9 of the ground state energy. In this narrow window (-4/9,0) there are infinitely many bound states.

\subsection{Singularity in Coulomb matrix elements}

In the next step we discuss singularity associated with direct electron-hole interaction for $\vec{k}=\vec{k'}$. Neglecting the summation over  G-vectors  for a moment, we assume constant exciton wavefunction inside $\delta k\times \delta k$ box centered around point ($k_x$,$k_y$), where $\delta k$ is defined in our single valley exciton theory as half of BZ area over number of k-points. These assumptions allow us to integrate $\vec{k}=\vec{k'}$ singularity analytically, with result:
\begin{equation}
\begin{split}
&\gamma\int_{k_x-\delta k/2}^{k_x+\delta k/2}\int_{k_y-\delta k/2}^{k_y+\delta k/2}dk_{x}'dk_{y}'\frac{A_{n}\left(\vec{k'}\right)}{\left|\vec{k}-\vec{k'}\right|}\approx \\
&A_{n}\left(\vec{k}\right)\gamma\int_{-\delta k/2}^{\delta k/2}\int_{-\delta k/2}^{\delta k/2}dk_{x}'dk_{y}'\frac{1}{\sqrt{k_{x}^{'2}+k_{y}^{'2}}}=\\
&A_{n}\left(\vec{k}\right)\gamma\left[2\ln{\frac{\sqrt{2}+1}{\sqrt{2}-1}}\right]\delta k = A_{n}\left(\vec{k}\right)\textrm{V}_{\textrm{sin.}}.
\end{split}
\label{eq23}
\end{equation}
We checked that $\vec{G} \neq 0$ contributions to singular term is ~2 orders of magnitude smaller than $\vec{G}=0$ and corrections from Rytova-Keldysh screening  (which are k' dependent) are negligible as well. Interaction form factors for singular terms are equal to 1.

\section{Computational details for single valley exciton calculations}

\subsection{Solution of Bethe-Salpeter equation with direct electron-hole interactions}

As a first step of computation of exciton spectrum in a single valley, we neglect a much weaker repulsive exchange interaction and retain only direct electron-hole attraction. We solve the following Bethe-Salpeter equation on a finite k-grid associated with one valley:
\begin{equation}
\begin{split}
&\left[\Delta E \left(\vec{k} \right) - \Delta_{\textrm{GAP}}- V_{\textrm{sin.}} \right]A_{n}\left(\vec{k} \right) \\
&- \sum_{\vec{k'}\neq\vec{k}}^{1/2\textrm{BZ}} \left(\delta k \right)^{2} V^{D}\left(\vec{k},\vec{k'}\right) A_{n}\left(\vec{k'}\right)=E_{n}A_{n}\left(\vec{k}\right),
\end{split}
\label{eq24}
\end{equation}
where diagonal singular correction, Eq. (\ref{eq23}), is taken as $V_{\textrm{sin.}}=3.53 \gamma\delta k$ , screened value of $V^{D}$ is taken and exciton energies $E_{n}$ are measured from the gap energy. We note that calculations of direct electron-hole interaction form factors are a bottleneck in our computations. However, contrary to the usual approximation in GW-BSE \cite{Deslipe_Louie_2012, Kammerlander_Attaccalite_2012} where matrix elements are computed on a sparse grid (e.g. $12\times 12 \times 1 = 144$ k-points\cite{Qiu_Louie_2013, Qiu_Louie_2013_erratum}) and interpolated in between, we do not interpolate our matrix elements and calculate them accurately on a dense grid, usually $\sim$7000 k-points per valley, unless stated otherwise. Further convergence details are presented in Appendix A and Appendix E.

\subsection{Computational details for calculations of exciton fine structure}

Here we discuss the spin splitting of conduction and valence band states and the effect of spin-dependent exchange interaction. The spin splitting of valence and conduction bands is included in electron-hole energy difference $\Delta E\left(\vec{k}\right)=\varepsilon_{\textrm{CB}}^{\sigma}\left(\vec{k}\right) - \varepsilon_{\textrm{VB}}^{\sigma'}\left(\vec{k}\right)$. Spin-dependent dispersion is calculated using spinfull Hamiltonian $H^{\textrm{TB}}_{\textrm{SOC}}=H^{\textrm{TB}}\otimes \hat{1}_{2\times 2}+\textrm{diag}\left[H_{\textrm{SOC}}(\sigma=1), H_{\textrm{SOC}}(\sigma=-1)\right]$, where $H_{\textrm{SOC}}(\sigma) = \textrm{diag}\left[-\sigma\lambda_{\textrm{Mo}},0,\sigma\lambda_{\textrm{Mo}},-\sigma\lambda_{\textrm{S}_{2}}/2,0,\sigma\lambda_{\textrm{S}_{2}}/2\right]$. To reproduce values of splitting obtained from \itshape ab initio \upshape we take for MoS$_{2}$ $\lambda_{\textrm{Mo}}=0.067 $ eV and $\lambda_{\textrm{S}_{2}}=0.02 $ eV.

This allows us, using Eq. (\ref{eq24}), to calculate exciton's fine structure (FS) in valley $+K$ and obtain A and B, bright and dark, excitonic levels. The time reversal symmetry implies that these levels are related between $+K$ and $-K$ valleys by $E_n(+K) = E_n(-K)$ and simultaneous spins flipping. In the  presented choice of gauge  we checked also that direct matrix elements have the following property $V\left(-\vec{k},-\vec{k}'\right)=V\left(\vec{k}',\vec{k}\right)$, therefore exciton wavefunctions between valleys are related by $A_{n}^{*}\left(-\vec{k}\right)=A_{n}\left(\vec{k}\right)$. 

To lower computational effort and clarify physics behind effects of exchange interaction, we write exciton's fine structure (FS) Hamiltonian in the basis of $n=1-4$ exciton states (first two shells). Energies of "Ising" excitons\cite{Guo_Fleming_2019} are written using matrix notation, e.g., for lowest bright A exciton, as $_{\uparrow}^{\uparrow}\hat{A}^{+K}_{\textrm{bright}}=\textrm{diag}\left(E_{1s},E_{2p_1},E_{2p_2},E_{2s}\right)$ . This allows us to write the full B.-S. problem as  
\begin{gather}
\hat{H}^{FS}=
\begin{bmatrix}
H_{+K,+K} & H_{+K,-K} \\
H_{+K,-K}^{\dagger}	& H_{-K,-K}  \\ 
   \end{bmatrix}
\label{eq25}
\end{gather}
\begin{gather*}
H_{+K,+K}=
\begin{bmatrix}
      _{\uparrow}^{\uparrow}\hat{A}^{+K}_{_{\textrm{bright}}}+V^{X} & 0 & V^{X} &0 \\
0 &     _{\uparrow}^{\downarrow}\hat{A}^{+K}_{_{\textrm{dark}}} & 0 & 0 \\
V^{X\dagger} & 0&     _{\downarrow}^{\downarrow}\hat{B}^{+K}_{_{\textrm{bright}}}+V^{X} &  0 \\ 
0& 0& 0& _{\downarrow}^{\uparrow}\hat{B}^{+K}_{_{\textrm{dark}}}
   \end{bmatrix}
\end{gather*}
\begin{gather*}
H_{-K,-K}=
\begin{bmatrix}
      _{\downarrow}^{\downarrow}\hat{A}^{-K}_{_{\textrm{bright}}}+V^{X} & 0 & V^{X} &0 \\
 0&     _{\downarrow}^{\uparrow}\hat{A}^{-K}_{_{\textrm{dark}}} & 0 & 0 \\
V^{X\dagger} &0 &     _{\uparrow}^{\uparrow}\hat{B}^{-K}_{_{\textrm{bright}}}+V^{X} &  0 \\ 
0& 0& 0& _{\uparrow}^{\downarrow}\hat{B}^{-K}_{_{\textrm{dark}}}
   \end{bmatrix}
\end{gather*}
\begin{gather*}
H_{+K,-K}=
\begin{bmatrix}
V^{X} & 0 & -V^{D}+V^{X} & 0\\
 0& 0 & 0 & -V^{D} \\
-V^{D}+V^{X} & 0&V^{X} &0  \\ 
0&-V^{D} &0 &0
   \end{bmatrix}
\end{gather*}
where $V^{D/X}$ are interactions "missing" in Eq. (\ref{eq24}). Using such approach and knowledge about unitary transformation diagonalizing sub-blocks of FS Hamiltonian, one can write intra-valley exchange interactions as a perturbation of "Ising" excitons 
\begin{equation}
\left[UV^{X}U^{\dagger} \right]_{ij}=\sum_{\vec{k},\vec{k'}}^{1/2 \textrm{ BZ}} A_{i}^{*}\left(\vec{k}\right) A_{j}\left(\vec{k'}\right)V^{X}\left(\vec{k},\vec{k'}\right).
\label{eq26}
\end{equation}
Here summations over exciton states can be greatly reduced due to localization of excitonic wavefunctions in k-space. We can now  diagonalize new FS Hamiltonian to obtain excitonic levels "corrected" by exchange interaction. We note that in general intra-valley interactions are stronger than inter-valley ones and major effect on FS comes from exchange interaction acting on A(B) bright exciton blocks (diagonal correction +$V^{X}$ in matrix in Eq. (\ref{eq25})).

\section{Numerical Results and Discussion}

\subsection{Role of $Q$ points and band nesting on exciton spectrum}

\begin{figure}[ht]
\centering
\includegraphics[width=0.5\textwidth]{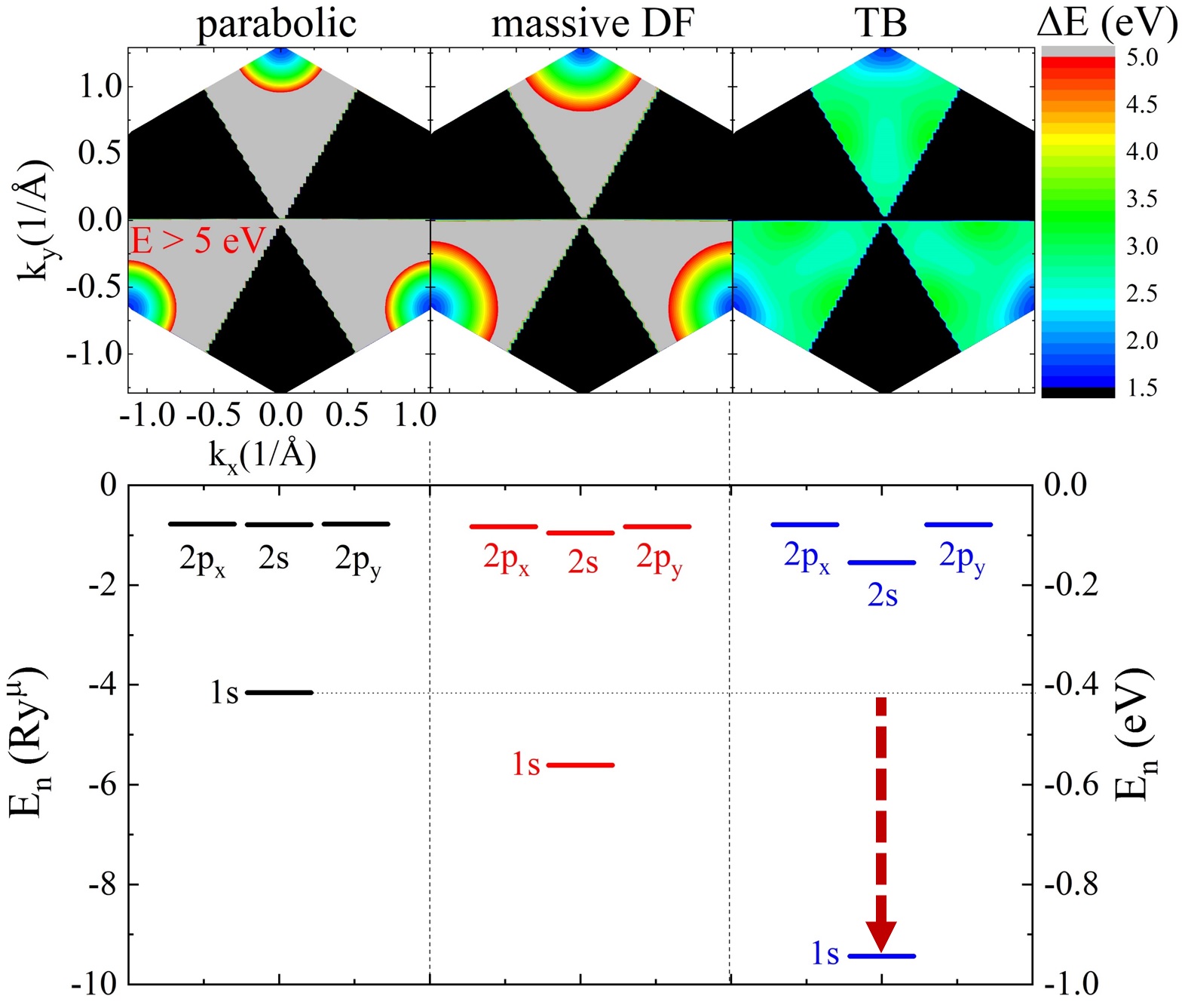}
\caption{Evolution of excitonic spectrum for three different models of dispersion, simplified $1/|q-q'|$ 2D interaction and static screening set to give effective excitonic Rydberg $\sim$100 meV for parabolic model dispersion. Red arrow points trend of excitonic spectrum (to be compared with arrow on Fig. \ref{fig4}).} 
\label{fig3}
\end{figure}

We start with describing our central result, the role of $Q$ points being a consequence of conduction and valence band mixing and resulting band nesting, on the exciton spectrum shown in Fig. \ref{fig3}. Using three models of electron-hole energy dispersion, namely \itshape ab initio \upshape based tight-binding, a simplified massive Dirac fermion model and a parabolic dispersion model around $K$ point, combined with  a simplified form of direct 2D statically screened electron-hole interaction,  $1/ |q-q'|$ ,  the  Bethe-Salpeter  equation, Eq. (\ref{eq24}), is solved. The parabolic dispersion of electron-hole pair corresponds to the ideal 2D exciton. Our numerical calculations reproduce analytical results, with the energy of the ground exciton $1s$ state 
 $E_{n=1} =-4$ Ry$^{\mu}$ (with  Ry$^{\mu}$ set to 100 meV ) and triply degenerate second shell with $2s$, $2p_{x}$ and $2p_{y}$ states, with energy $E_{n=2-4} = -4/9$ Ry$^{\mu}$. We next keep the same interaction, but change electron-hole dispersion to massive Dirac fermion model. We find increased binding energy of 1s state, which can be explained by larger effective mass averaged over valley due to linear, instead of parabolic, electron-hole dispersion away from the K-point (see Fig. \ref{fig2}(b)). Switching to tight-binding dispersion, which turns on contribution from three $Q$ points, magnifies this effect further, changing the binding energy of the lowest 1s state up to almost -10 Ry$^{\mu}$ and significantly renormalizes the 1s-2s exciton energy levels spacing. The effect of $Q$ points on exciton wavefunctions is analyzed in details in Appendix C.

\subsection{The role of screening on the exciton spectrum}

\begin{figure}[ht]
\centering
\includegraphics[width=0.5\textwidth]{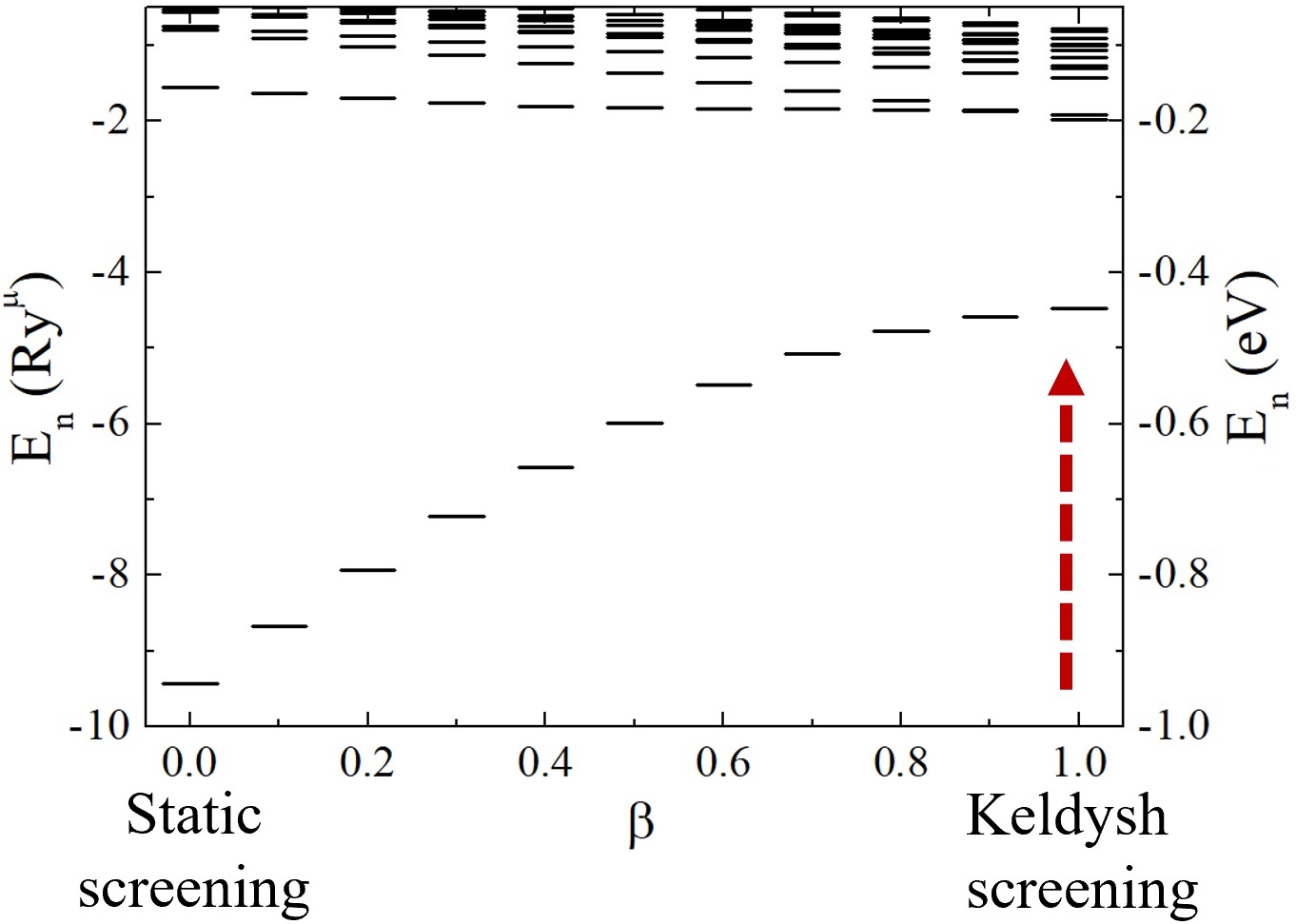}
\caption{Dependence of excitonic spectrum on screening model, showing transition from static screening ($\beta=0$) to Rytova-Keldysh screening ($\beta=1$). Dispersion model is taken as parabolic, electron-hole interaction is still simplified to $1/|q-q'|$ and parameter $\alpha=2.2$ \AA.} 
\label{fig4}
\end{figure}
It is known that screening by 2D systems is not well represented by static, homogeneous dielectric constant $\epsilon_{r}^{stat.}$ \cite{Rytova_1967,Keldysh_1979}, but requires non-local effects. To show how the non-local screening influences the exciton excited states, we write direct matrix element entering Eq. (\ref{eq24}) in the following form: 
\begin{equation}
\begin{split}
&V^{D}\left(\vec{k},\vec{k'}\right) =\gamma\sum_{\vec{G}}\frac{F^{D}\left(\vec{k},\vec{k}',\vec{G}\right)}{\left|\vec{k}'-\vec{k}-\vec{G} \right|}   \times\\
& \left(\frac{1-\beta}{\varepsilon_{r}^{\textrm{stat.}}} + \frac{\beta}{\varepsilon_{r}^{\textrm{R-K}}\left(1+2\pi \alpha \left|\vec{k'}-\vec{k}-\vec{G} \right| \right) }\right),
\end{split}
\label{eq27}
\end{equation}
where $\alpha$ is electron polarizability, treated as a parameter here, and $\beta$ controls the transition from homogeneous  to non-local screening.

In Figure \ref{fig4} we show how switching between static ($\beta=0$) and Rytova-Keldysh  ($\beta=1, \alpha=2.2$ \AA) form of screening affects the excitonic spectrum. These results were obtained for tight-binding dispersion and simplified 2D $1/|q-q'|$ direct electron-hole interaction. One can observe strong renormalization of exciton spectrum by different forms  of screened interactions.  The renormalization  of $1s-2s$ energy separation can be observed, along with shifting the order of degenerate $2p_{x}$, $2p_{y}$ states with respect to the $2s$ state. The observed effects are  opposite to the effects shown in Fig. \ref{fig3}.

\subsection{The renormalization of X spectrum from "2D-like"  to "3D-like" }

\begin{figure}[ht]
\centering
\includegraphics[width=0.5\textwidth]{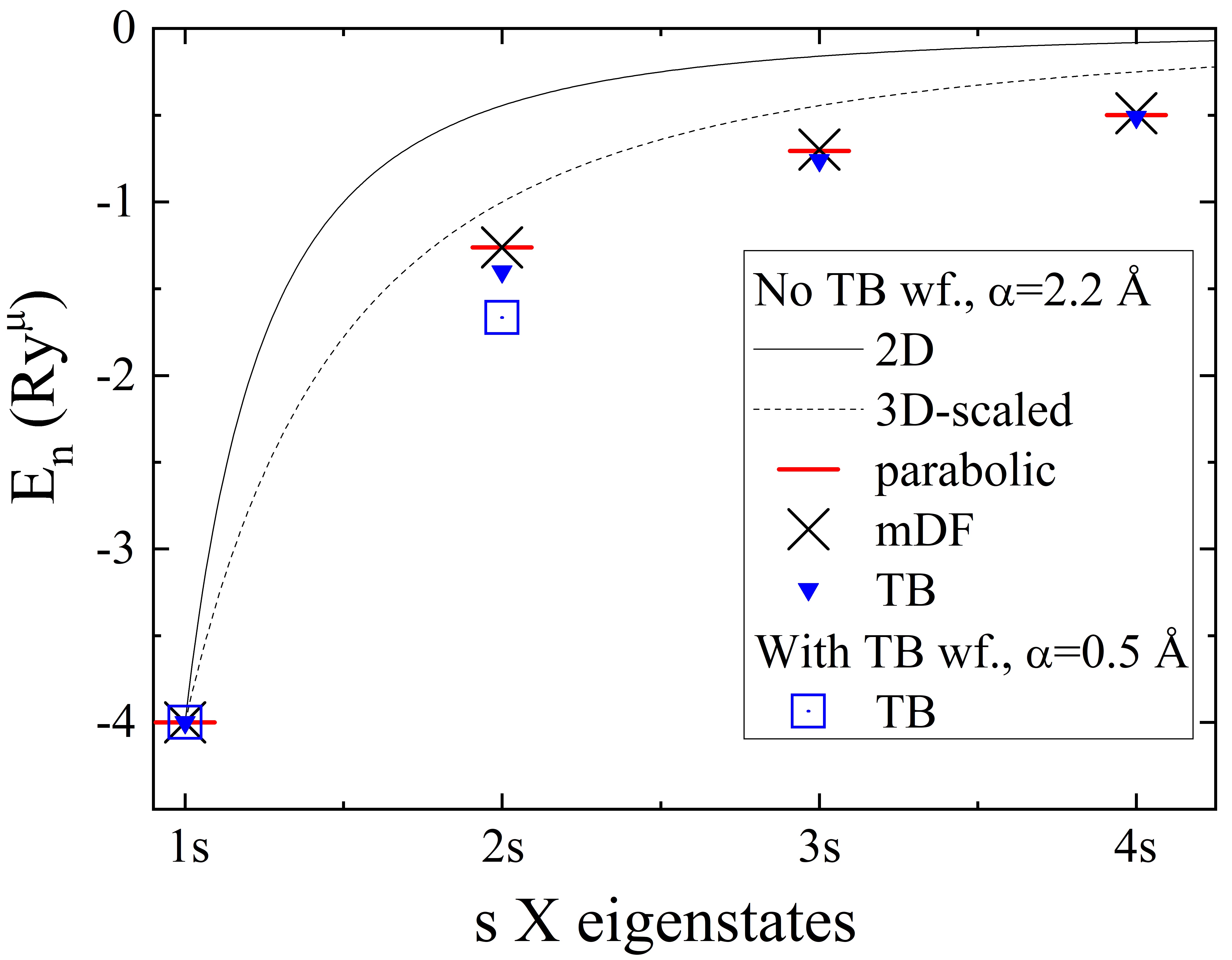}
\caption{Comparison between analytical results for 3D and 2D exciton hydrogen-like s-excitons series and our numerical results for different dispersion (parabolic effective mass, massive Dirac fermion (mDF) and tight-binding (TB)) and interaction models (with/without tight-binding wavefunction effects) with Rytova-Keldysh-like screening. } 
\label{fig5}
\end{figure}

Here we discuss the combined effects of electron-hole dispersion and screening on the exciton spectrum. We were able to extract numerically the $1s$ to $4s$ exciton levels, as shown in Fig. \ref{fig5}. We conclude that  the energy of optically active s exciton levels renormalizes from the  2D Rydberg exciton series, solid line on Fig. \ref{fig5},  towards more "3D-like" series (dotted line), when all energies are scaled to the same $1s$ binding energy of
 $-4$ Ry$^{\mu}$. We note that  the accuracy of 
higher states  decreases due to the finite k-grid. 

The effect of  renormalization of the exciton spectrum has to be  attributed not only to non-local screening, but also to electron-hole dispersion model taking into account $Q$ points, as well as the effect of direct electron-hole Coulomb interaction form factor F. The inclusion of wavefunctions in the evaluation of Coulomb matrix elements shifts the energy levels, as shown by blue rectangles in Fig. \ref{fig5}. We were able to resolve only $1s$ and $2s$ states in this case. All calculations were performed on the same k-grid for consistency, with $\sim$7000 k-points per valley.

\subsection{Effects of form-factor and topology on electron-hole  interaction}

In the next step we discuss the effect of topology on direct electron-hole interaction. When only absolute value of form factor $\left|F^{D}\left(\vec{k},\vec{k'},\vec{G}\right)\right|$ from Eq. (\ref{eq14}) is taken when solving Bethe-Salpeter equation, we obtain degenerate $2p_{x}$ and $2p_{y}$ states in exciton spectrum. Turning on microscopic phases originating from valence and conduction band states  in those form factors, we find two mixed  $2p_{\pm 1}$ states :  $2p_{\pm 1}=2p_{x}\pm i2p_{y}$ , split in 
\begin{figure}[ht]
\centering
\includegraphics[width=0.5\textwidth]{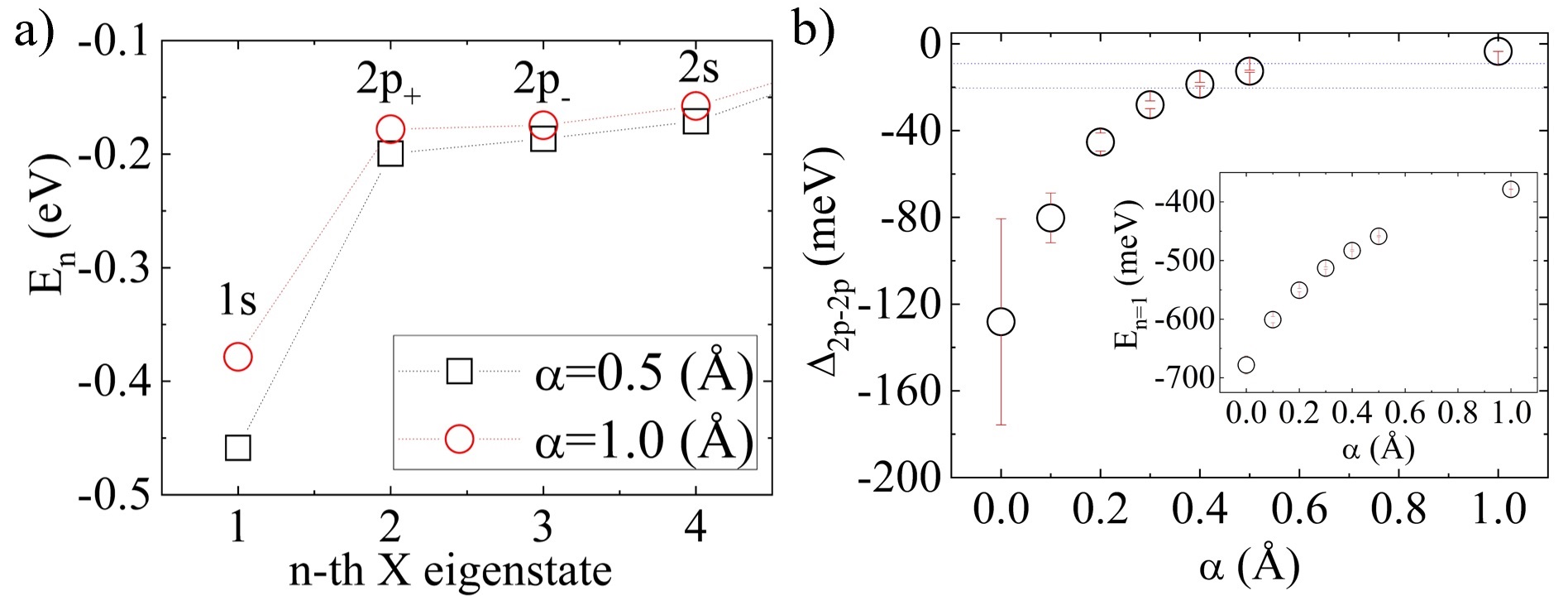}
\caption{(a) Exciton spectrum in MoS$_{2}$ taking realistic tight-binding dispersion and chiral interaction for two different Rytova-Keldysh screening parameters $\alpha$. (b) Dependence of topological $\Delta_{2p-2p}$ splitting (inset: $1s$ ground state binding energy) on Rytova-Keldysh screening parameter. Error bar shows estimated difference between summation over $G_{\textrm{cutoff}} = \left|G_1\right|$ reciprocal lattice vector and simplified single-G model (as described in main text). } 
\label{fig6}
\end{figure}
energy. Difference between k-space shape of $p_{x,y}$ exciton states versus $p_{\pm 1}$ is presented in Appendix B. This splitting, $\Delta_{2p-2p}$, depends heavily on screening. For example, for $\alpha=1.0$ we obtain 
$\Delta_{2p-2p}=3.5$ meV (the corresponding $1s$ state binding energy is then 378 meV)  and for $\alpha=0.5$ we get 
$\Delta_{2p-2p}=13.0$ meV ( the corresponding $1s$ state binding energy is 458 meV), see Fig. \ref{fig6}(a). 
The full dependence of $\Delta_{2p-2p}$ splitting on polarizability $\alpha \le 0.5$ is given in Fig. \ref{fig6}(b). The inset in  Fig. \ref{fig6}(b) shows the corresponding binding energies of $1s$ states. The error bars show estimated error one gets by neglecting the summation over $\vec{G}$ vectors in Eq. (\ref{eq14}), instead taking only one $\vec{G}$ vector such that $\left|\vec{k'}-\vec{k}-\vec{G}\right|$ is smallest on 1st BZ. We note that when screening is small, for strongly bound states the magnitude of splitting sensitively depends on details of the electron-hole interaction. 

Similar values for excitonic 2p shell  splitting were also reported using the simplified massive Dirac fermion model of interaction \cite{Wu_MacDonald_2015, Srivastava_Imamoglu_2015}. In this model the conduction to valence band coupling is linear in momentum.  We note that using such simplified Hamiltonian and eigenstates in the electron-hole form factor  is only valid close to the K - point. The proper mDF model Hamiltonian, correct for the whole first BZ (compare with Eq. (\ref{eq4})),  reads:
\begin{gather}
\hat{H}^{mDF}\left(\vec{k}\right)=
\begin{bmatrix}
\Delta/2 & g_{k}e^{i\theta_{k}}  \\
g_{k}^{*}e^{-i\theta_{k}} & -\Delta/2 
\end{bmatrix},
\end{gather}
with $g_{k}e^{i\theta_{k}}=t\exp(-i\vec{k}\vec{b})(1+\exp(i\vec{k}\vec{a_{2}})+\exp(i\vec{k}(\vec{a}_{1}+\vec{a}_{2})))$, $\vec{b}=\left(d_{\parallel},0\right)$ and $3/2 d_{\parallel}t=\hbar v_{\textrm{F}}$. The form factor $\Phi(k,k')$ of direct electron-hole interaction is then given by
\begin{equation}
\begin{split}
&\Phi(k,k')=\bigg[\sin\frac{\varphi_{k'}}{2}\sin\frac{\varphi_k}{2}\exp\left[-i\left(\theta_{k}-\theta_{k'}\right)\right]+ \\
&\qquad \qquad \cos\frac{\varphi_{k'}}{2}\cos\frac{\varphi_{k'}}{2}\bigg]^{2},
\end{split}
\end{equation}
with $\cos\varphi_{k}=\frac{\Delta/2}{\sqrt{\Delta^{2}/4+g_{k}^{2}}}$. When this model interaction and mDF dispersion is used, we obtain values of $2p-2p$ shell splitting of the order of $\approx$ 20 $\mu eV$, significantly lower than those from microscopic TB results. Only unphysical extension, assuming $g_{k}e^{i\theta_{k}}\approx\hbar v_{\textrm{F}} (iq_{x}-q_{y})$, with $\vec{q}$ measured from $\vec{K}$, increases the value of 2p-2p splitting to $\approx$ 33 meV. This value is further reduced by Rytova-Keldysh screening. We conclude here that the exciton spectrum obtained in the massive Dirac fermion model valid close to the bottom of the $K$ valley gives only a qualitative understanding of some aspects of the exciton spectrum.

\subsection{Effect of spin-orbit coupling in conduction band on the exciton spectrum}

\begin{figure}[ht]
\centering
\includegraphics[width=0.5\textwidth]{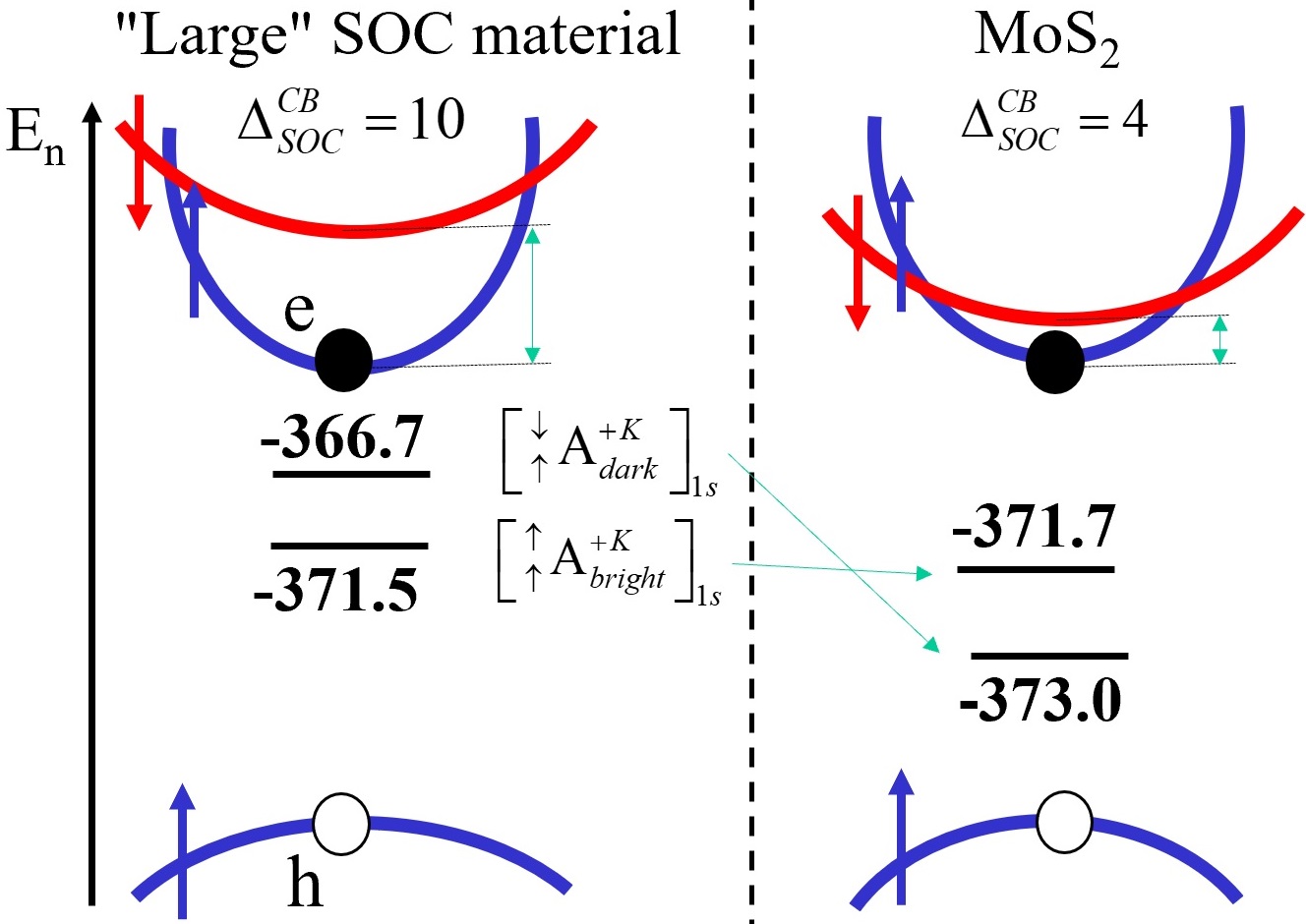}
\caption{(left) Spin ordering of bands in "bright" MX$_{2}$ material for large VB spin-splitting (lower band not shown) and 10 meV spin splitting in the CB. Black bars denote positions of dark and bright $1s$ A-exciton levels resulting solely from the interplay of spin-splitting and different effective masses of spin split bands, neglecting electron-hole exchange. (right) Ground state of exciton changes to dark one even for "bright" arrangement of spinfull CB bands due to the difference of effective masses.  All energies on the Figure are given in meV. Parameters here correspond to results presented on Fig. \ref{fig6} for $\alpha=0.5$ \AA. } 
\label{fig7}
\end{figure}

The exciton fine structure is determined by valleys, spin splitting of bands in each valley due to spin-orbit coupling (SOC) and by intra and inter-valley exchange interactions. We first focus on the SOC effect, which in TMDCs is equivalent to SOC induced Zeeman splitting. The splitting of CB and VB leads to  A and B excitons, with splitting  determined primarily by large spin-splitting in the VB, $\sim 140$ meV in MoS$_{2}$ \cite{Kadantsev_Hawrylak_2012}, resulting in A-B exciton splitting of 
$\sim125$ meV for $\alpha=0.5$ series on Fig. \ref{fig6}(a). A more subtle effect is connected with the A exciton dark-bright splitting, which is controlled by spin splitting in the CB. 
We start with a  material with "large" SOC splitting, $\sim 10$ meV,  with band ordering yielding bright lowest energy VB to CB transition, and with calculated $A_{\textrm{bright}}$ excitonic state below the dark one, as shown in Fig. \ref{fig7} left panel. 
However, the value of this dark-bright exciton splitting is not the same as spin splitting of conduction bands at the K-point. The large spin splitting  in the valence band and the coupling of VB and CB results in different, spin dependent,  CB effective masses and hence in different binding energy for exciton built with spin up/spin-down conduction band states.  Therefore, for low CB spin splitting as obtained in \itshape ab initio \upshape calculations,
$\Delta^{CB}_{SOC}=4$ meV in MoS$_2$ \cite{Kadantsev_Hawrylak_2012}, even though spin ordering of lowest energy single particle electron-hole transitions renders them still bright, the lowest excitonic state is dark, as illustrated in Fig. \ref{fig7} left and suggested also by GW-BSE calculations \cite{Qiu_Louie_2015, Torche_Bester_2019} and some experiments \cite{Molas_Potemski_2017}.

\subsection{The role of electron-hole exchange interaction on exciton’s fine structure}

\begin{figure}[ht]
\centering
\includegraphics[width=0.5\textwidth]{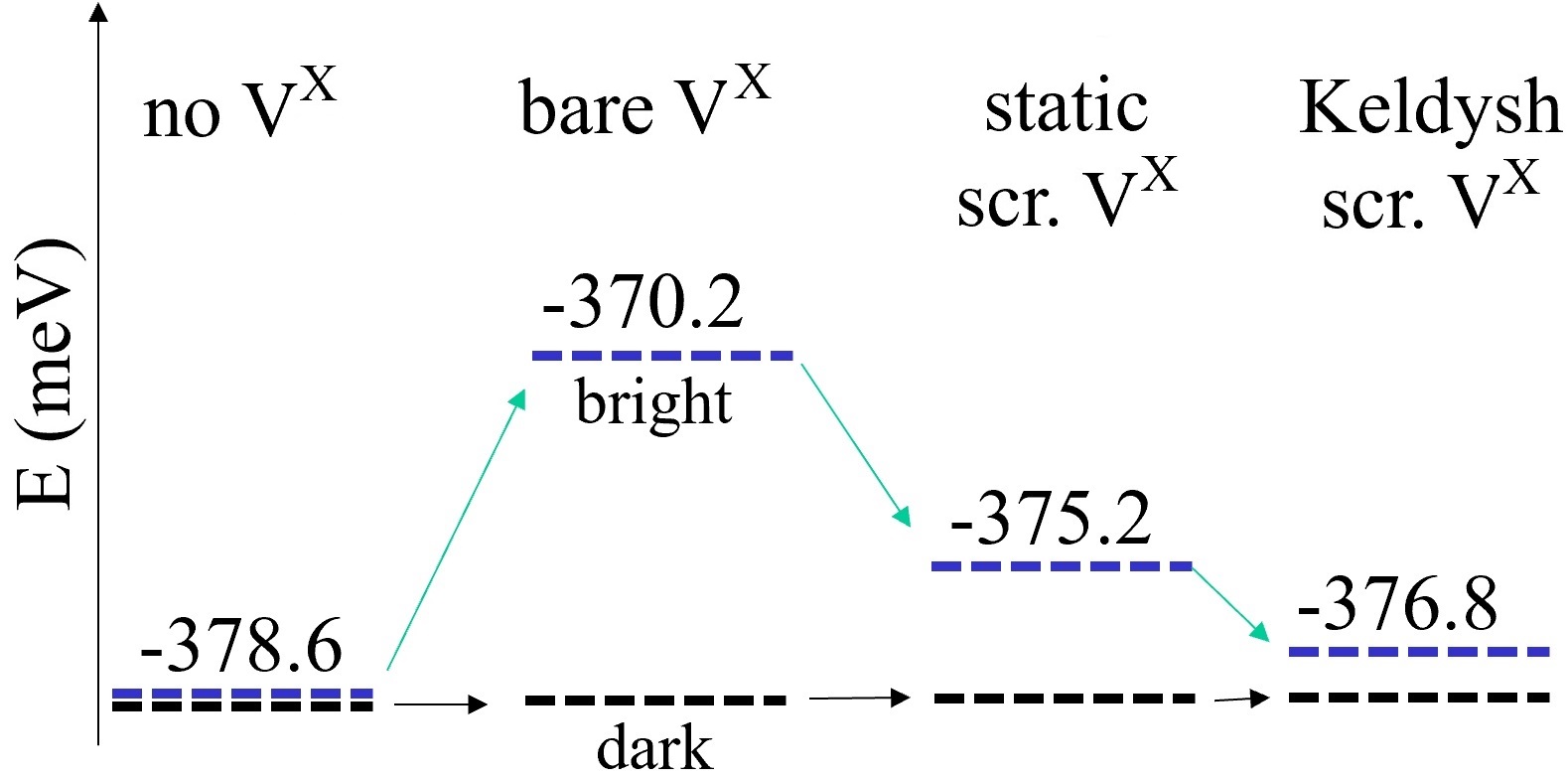}
\caption{The effect of electron-hole exchange interaction and screening under the assumption of spin degenerate CB.  All values shown are given in meV. Significant dark-bright splitting due to unscreened electron-hole exchange interaction (bare $V^X$) is reduced due to statically screened exchange, and can further be lowered by addition of Rytova-Keldysh-like screening. In all cases exciton ground state is dark. } 
\label{fig8}
\end{figure}
We now turn to discuss the role of intra-valley exchange interaction. Turning off the effect of SOC and neglecting exchange results  in  bright and dark exciton states being degenerate, as illustrated on the "no $V^{X}$" panel in Fig. \ref{fig8}. Turning on repulsive exchange matrix elements  with different models of screened interaction, from bare to static to Rytova-Keldysh, we show in Fig. \ref{fig8} that due to effective electron-hole repulsion, the bright excitonic state becomes less bound and its energy increases. At the same time the dark exciton state is unaffected by exchange interaction and hence becomes the lower energy state.
The effect of exchange follows the same trend as the effect of  SOC, with discussion above.  We note that the magnitude of the dark-bright exciton splitting depends on the  exchange matrix elements, evaluated here with the same screened interaction as enters the direct electron-hole attraction. Green’s function based DFT+GW+Bethe-Salpeter approach \cite{Qiu_Louie_2015} suggests taking unscreened value of exchange interaction. This issue needs further investigation \cite{Torche_Bester_2019, Benedict_2002}. However, irrespective of the approach, we conclude that the dark A exciton  is the lowest energy exciton state, $\sim 3-10$ meV below the bright exciton
 in MoS$_{2}$.

\subsection{Inter-valley exciton scattering}

We discuss here two valley excitons with zero total momentum $Q$. Following the argument based on C$_{3}$ symmetry \cite{Yu_Yao_2014, Qiu_Louie_2015} the intervalley exchange coupling of Q=0  excitons in $+K$ and $-K$ valleys should vanish and the exciton spectrum should be degenerate. Hence numerical calculation of intervalley exchange of  $Q=0$ center-of-mass momentum  excitons is a sensitive test of numerical accuracy. We note that in our calculations of intervalley exchange, keeping the same convergence parameters in both direct and exchange interactions, results in small inter-valley exchange coupling, resulting in $<1$ meV 1s-1s exciton splitting. However, the coupling decreases  to zero with increasing accuracy  of calculation, as expected from symmetry arguments  \cite{Yu_Yao_2014, Qiu_Louie_2015}.

\section{Conclusions}

We presented here a theory of excitons in monolayer MoS$_2$ starting from the atomistic \itshape ab initio \upshape based tight-binding model. We discussed to what extent the exciton spectrum reflects the approximate models of MoS$_2$, from free electrons and holes in a 2D semiconductor, to excitons described by a massive Dirac fermion model with topological moments and Rytova-Keldysh screening, to a model capturing band nesting and three $Q$ points per Dirac fermion. We constructed a theory of single valley exciton and formally built different levels of approximations of electron-hole dispersion, interaction and screening, studying these different contributions  separately. The effect of $Q$ points on excitonic spectrum, together with Rytova-Keldysh screening, was shown to produce transition from standard 2D exciton Rydberg series towards more "3D-like" excitonic series of s - like states, consistent with experiments. Then the effect of chirality of direct electron-hole interaction was analyzed, showing how "topological" splitting of $2p_{\pm}$ states already present in massive Dirac fermion model is modified by screening and existence of $Q$ points. The inclusion of spin-orbit coupling and electron-hole exchange interaction was found to affect exciton fine structure, showing that even for "bright" ordering of spin-polarized conduction bands one obtains dark excitonic ground state, and dark-bright splitting is further magnified by exchange interaction, resulting in MoS$_{2}$ being optically dark material. 

\section*{Acknowledgments}
M.B. and P.H. thank Y. Saleem, M. Cygorek, P.K. Lo, S.J. Cheng, M. Korkusinski, J. Jadczak, L. Bryja, P. Potasz and A. Wojs for discussions.  M.B., L.S., and P.H. acknowledge support from NSERC Discovery and QC2DM Strategic Project grants as well as uOttawa Research Chair in Quantum Theory of Materials, Nanostructures and Devices. M.B. acknowledges financial support from National Science Center (NCN), Poland, grant Maestro No. 2014/14/A/ST3/00654. Computing resources from Compute Canada and Wroclaw Center for Networking and Supercomputing are gratefully acknowledged.


\section*{Appendix A: A grid of k-points and convergence of excitonic spectrum}

\begin{figure}
\centering
\includegraphics[width=0.5\textwidth]{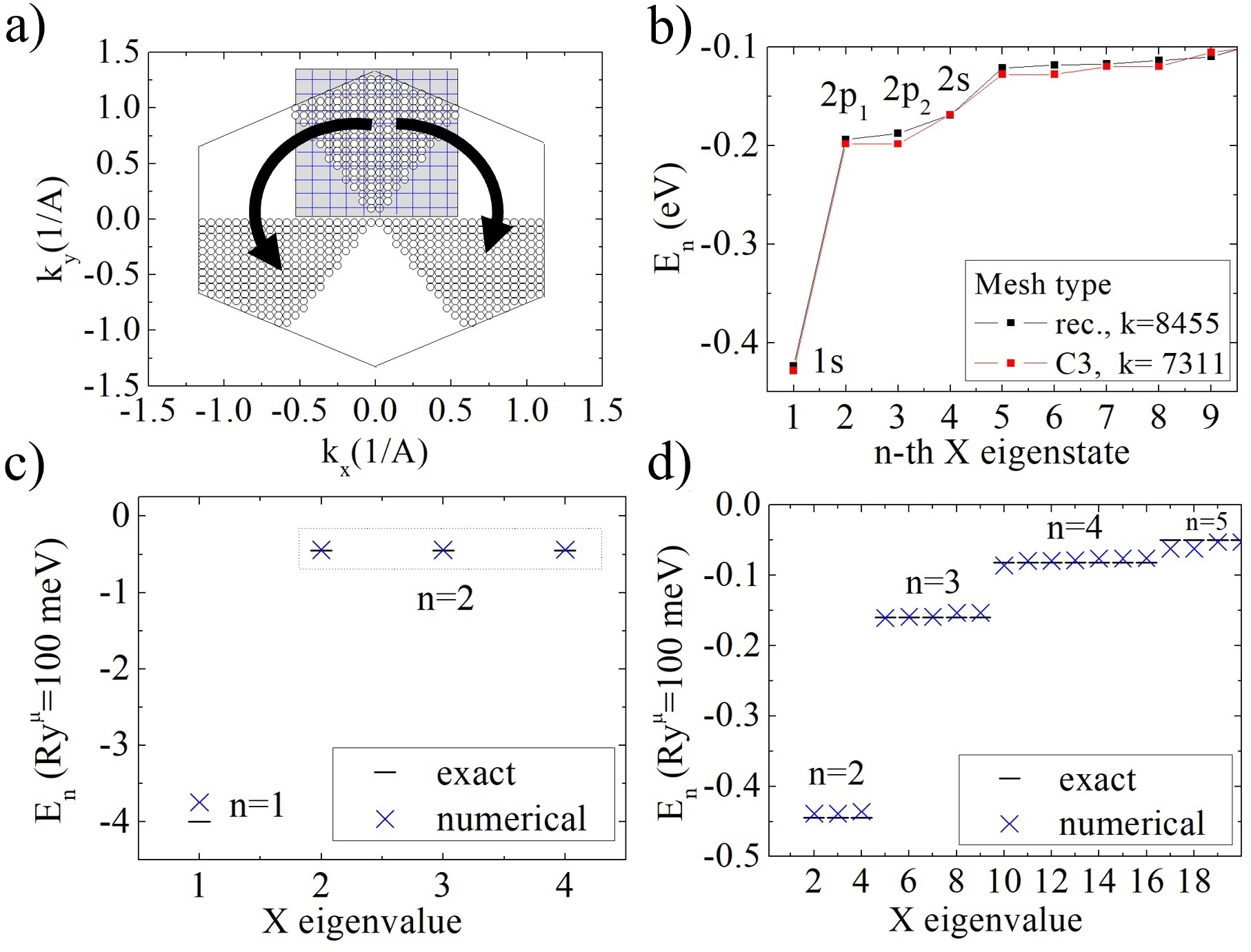}
\caption{(a) Schematic way of choosing C$_3$ symmetric k-point coordinates used in numerical calculations.  (b) Exciton spectrum comparing C$_3$ symmetric and strictly rectangular grid of k-points. (c) Convergence of numerical results to analytical ones with respect to very high density of C$_{3}$ k-point grid (120 000 k-points, TB electron-hole energies, simplified $1/|q-q'|$ interaction, static screening).  (d) Convergence of excited exciton states up to n=4 shell. In (b-d) cases hydrogen-like 2D exciton is studied (parabolic electron-hole energies, simplified $1/|q-q'|$ interaction, static screening).}
\label{figA1a}
\end{figure}

The numerical solution of Bethe-Salpeter equation,  Eq. (\ref{eq8}), has to be carried out on a lattice of k-points, which needs to discretize the half of the Brillouin zone associated with +K valley, as shown in Fig. \ref{figA1a}(a). We note that  this discretization can be performed in several different ways. We start with "brute - force" rectangular discretization, in which first the full hexagonal BZ is divided into small regions and then k-points in their centers  are ascribed to a given valley. In Fig. \ref{figA1a}(b) (black line) we show that this scheme breaks the $C_3$ symmetry and therefore lifts the degeneracy of the 2p states, which should be present for TB 
dispersion  and real electron-hole interaction which neglects topological effects. To restore this degeneracy, we apply a slightly modified scheme, in which we first select k-points in one "kite" of points in one valley of the BZ and then rotate this kite by $\pm2\pi/3$. This procedure restores expected degeneracy of 2p states in numerical calculations. We note also that a grid of $\sim 7000$ k-points  is sufficient to describe the  n=1 and n=2 shells, however it fails to accurately describe $n>2$ excited shells. To show precisely how modifications of electron-hole dispersion affect the exciton  s-series, we perform large-scale calculation for a lattice of $1.2\cdot 10^5$ k-points. We solve Eq. (\ref{eq20}) numerically using equivalent Eq. (\ref{eq8}), checking that it gives correct shell energies and degeneracies up to n=4 shell. In Fig. \ref{figA1a}(c-d) we compare analytical and numerical solutions for n=1-2 and n=2-5, respectively,  and show perfect agreement up to n=4 shell.

\section*{Appendix B: Exciton wavefunctions in k-space}

\begin{figure}
\centering
\includegraphics[width=0.5\textwidth]{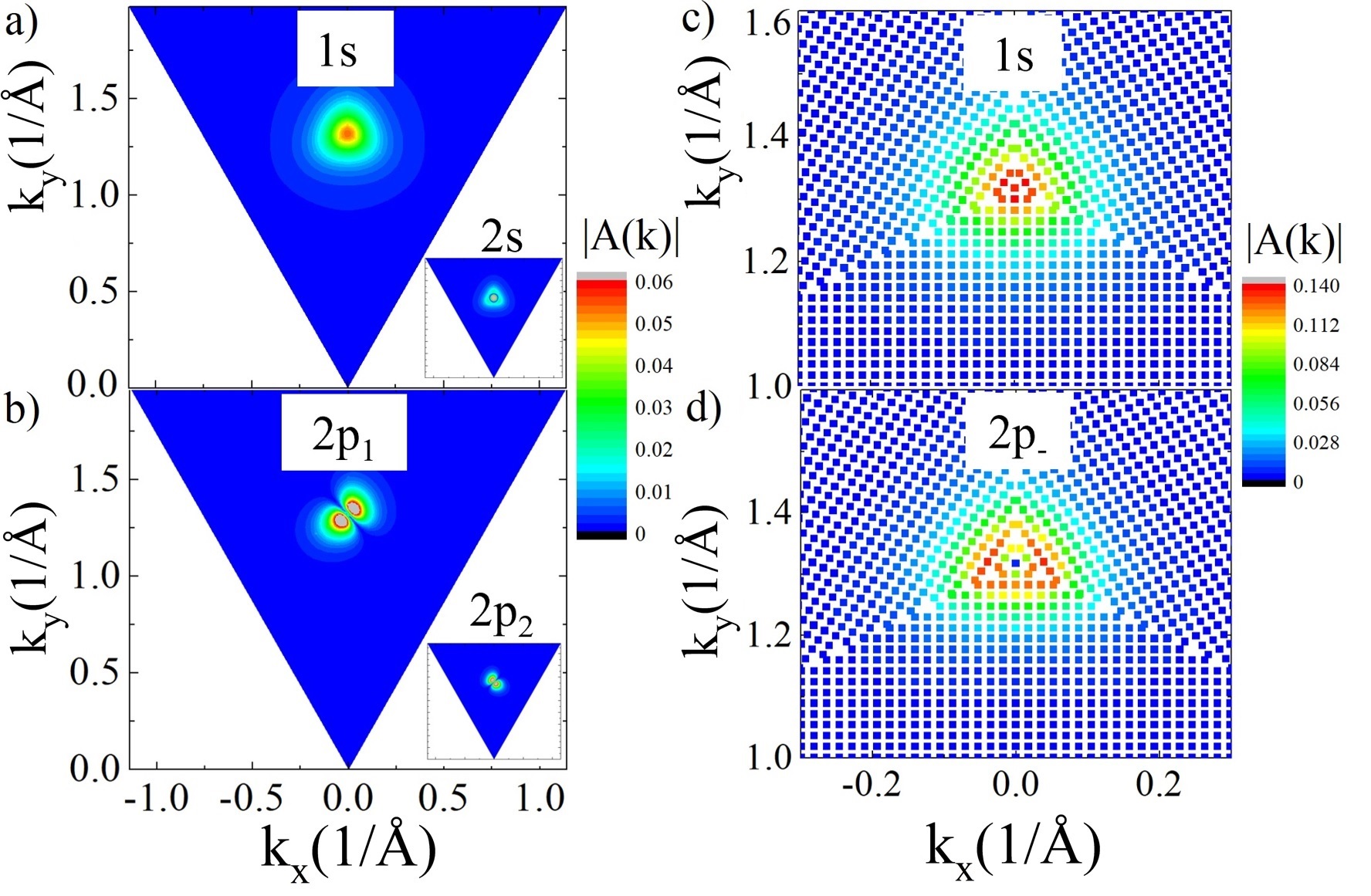}
\caption{: (a-b) Exciton wavefunction amplitudes $\left|A_{n}\left(\vec{k}\right)\right|$ in  K valley calculated using tight-binding dispersion and simplified $1/|q-q'|$ interaction. (a) shows $1s$ type state (inset: $2s$ state) and (b) presents $2p_{1}$ state (inset: $2p_{2}$), as described on Fig.\ref{figA1a}(c). (c-d) Corresponding results for full tight-binding model (dispersion and chiral electron-hole interaction), showing different character of topologically "mixed" $2p_{-}$ state ($1s$ and 
$2p_{-}$ correspond to Fig.\ref{fig6}(a)).} 
\label{figA2}
\end{figure}

Exciton wavefunctions and exciton binding energies describe the excitonic spectrum. For real  electron-hole interaction 
 they are real functions due to real, Hermitian, eigenvalue problem defined by Eq. (\ref{eq8}). By plotting them on triangular k-grid representing single valley, we are able to visualize the  s-, p- and d- symmetries of excitonic states, as long as exciton Bohr radius is reasonably larger than k-point spacing. Examples of 1s (2s in inset) and $2p_1$ ($2p_2$ in inset) excitonic states are shown in Fig. \ref{figA2}(a) and Fig. \ref{figA2}(b), respectively. One can clearly distinguish s-like states having maximum at $K$ point (center of triangle) and p-like states having minimum at the center of the valley. We note that when 1s state is plotted on logarithmic scale, excitonic function is so large in parabolic model with static screening and simplified electron-hole interaction, that it "touches" the boundaries of k-grid defining K-valley, which explains slight difference of analytical (infinite k- space) and numerical (restricted to 1 valley in the BZ) solutions, as shown for first exciton eigenvalue on Fig.\ref{figA1a}(c).

Turning on chirality (complex character)  of electron-hole Coulomb direct interaction results in complex wavefunctions of all exciton states. To study rotation of 1s state in $+K$ and $-K$ valley we calculate independently two sets of matrix elements for all $k \to k'$ exciton scatterings in both valleys. Exciton eigenergies in +K (-K) valley from those two separate calculations do not differ more than $1\%$. Studying phase of those 1s excitons (module plotted on Fig. \ref{figA2}(c) ) around $\pm$K-points we conclude that it rotates around valley minimum by $\pm 6\pi$, numerically proving  symmetry of wavefunctions between the two nonequivalent valleys $A_n(k)=A^{*}_{n}(-k)$ expected from Eq. (\ref{eq8}). We note that 2p states also become complex. Furthermore, they become a mixture of real $p_x$- and $p_y$- like state, forming topologically split $2p_{\pm} = 2p_x \pm i 2p_y$ states. When plotting the module of these states, overall symmetry is naturally circular, as for s-like states. Only the minimum in the center of the valley allows us to distinguish  between s-like and p-like symmetry, see Figure \ref{figA2}(d).

\section*{Appendix C: Detailed role of $Q$ points on exciton spectrum}

\begin{figure}
\centering
\includegraphics[width=0.5\textwidth]{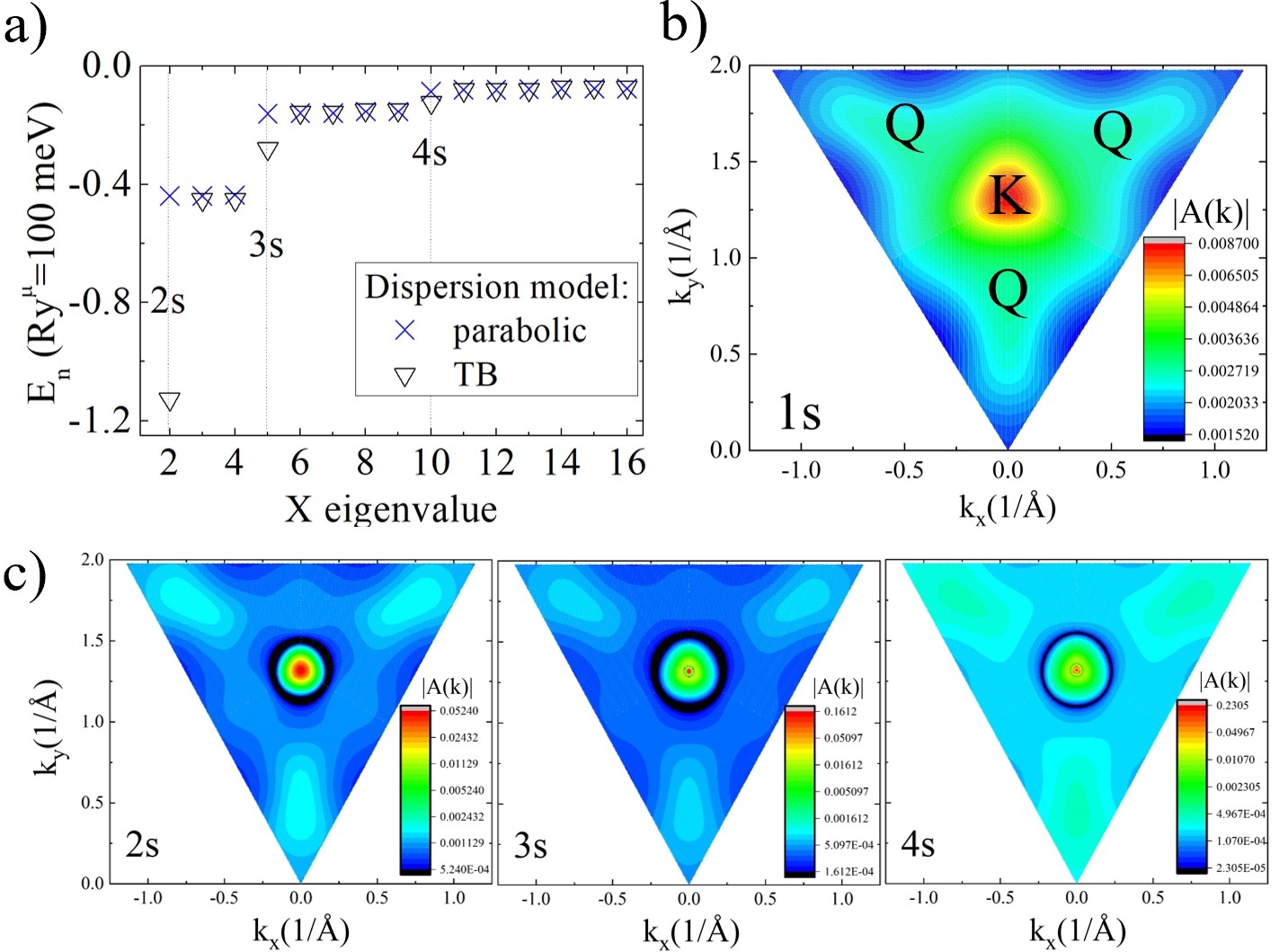}
\caption{(a) Spectrum of excited exciton states for parabolic and TB dispersions (static screening, no form factors in interaction). (b) 1s, (c) 2s-4s exciton wavefunctions on log scale, showing how three $Q$ points break rotational symmetry of s-states.}
\label{figA1b}
\end{figure}

Using dense grid of  k-points, as described in Appendix A, allows us to determine the effect of $Q$ points on excited exciton states. The renormalization of 2s to 4s states is presented in Fig. \ref{figA1b}(a), along with the effect of $Q$ points on 1s to 4s exciton wavefunctions, plotted here on logarithmic scale (Fig. \ref{figA1b}(b-c)), for module of the exciton wavefunction $|A_{n}|$. We note that p-like states from n=2 shell also become deformed due to existence of three $Q$ points. Interestingly, this effect is asymmetric due to modification of two-node function (p-like state) by $C_3$ symmetric "background" from three $Q$ points, but simultaneously it leads to significantly weaker renormalization, as shown in Fig. \ref{figA1b}(a).

\section*{Appendix D: Product densities of Bloch wavefunctions}

\begin{figure}
\centering
\includegraphics[width=0.5\textwidth]{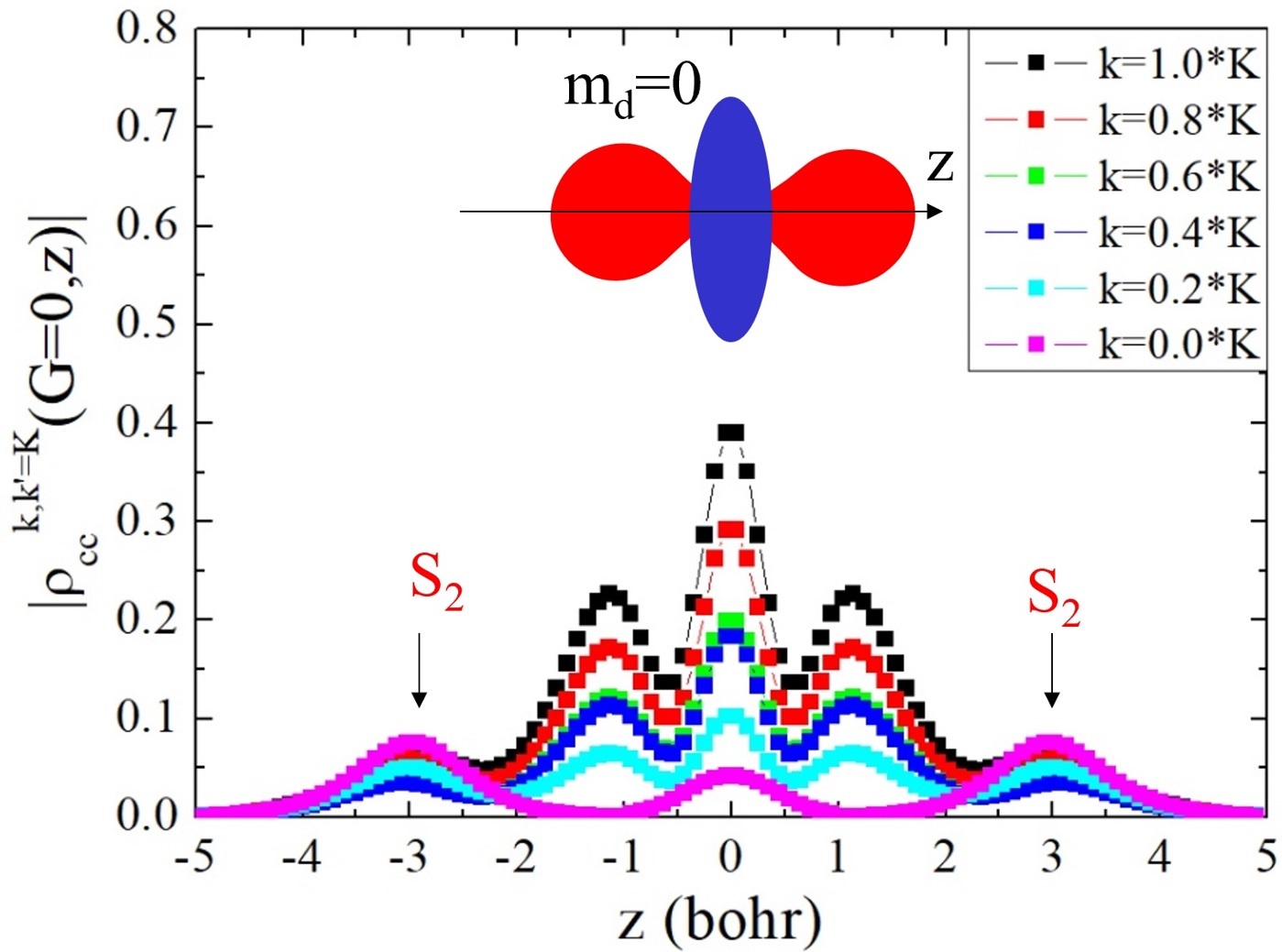}
\caption{Example of dependence of conduction band Bloch product densities $|\rho_{\textrm{cc}}|$ for $\vec{G}=0$ and k' set to K-point for varying k along the K-$\Gamma$ line in the BZ. Inset schematically shows $m_{d}=0$ orbital contributing mostly to conduction band, which combined with arrows denoting the sulfur dimer positions in the  z direction, allow to understand qualitatively the structure of z dependence of product density functions.} 
\label{figA3}
\end{figure}

The fundamental objects which  affect  the form factors of both direct and exchange electron-hole Coulomb interaction are pair densities (also called co-densities) constructed from eigenvectors of the tight-binding Hamiltonian and  Bloch wavefunctions, as explained in Eq. (\ref{eq16}). In general, these are complicated objects, depending on two wavevectors (k and k'), band indices and reciprocal lattice vectors $\vec{G}$. However, some of their features can be understood, as shown in Fig. \ref{figA3}. We present the z-dependence of a class of pair densities in the conduction band, $\rho_{cc}^{kk'}(z)$, showing their z-dependence, which can be understood in terms of integrating the 2D planes corresponding to a given z,  weighted by orbital contribution for a given band and k vector. For example, in conduction band around $K$ point, both Mo $m_d=0$ and sulfur p-orbitals contribute to the bandstructure, which is reflected in general structure of corresponding co-density. $\rho_{cc}$ reflects the nodal structure of d orbital, with additional contribution from sulfur atoms, positioned symmetrically away from the z=0 plane. For this particular type of co-density function, when we go away with one of the k-points from valley center at K, we notice overall decrease of $\rho_{cc}$, affecting stronger Mo contribution (around z=0) than sulfur dimer ($S_{2}$) contributions, centered around z positions of sulfur atoms ($\approx \pm 3 $ bohr).

\section*{Appendix E: Convergence of tight-binding Coulomb matrix elements}

\begin{figure}
\centering
\includegraphics[width=0.5\textwidth]{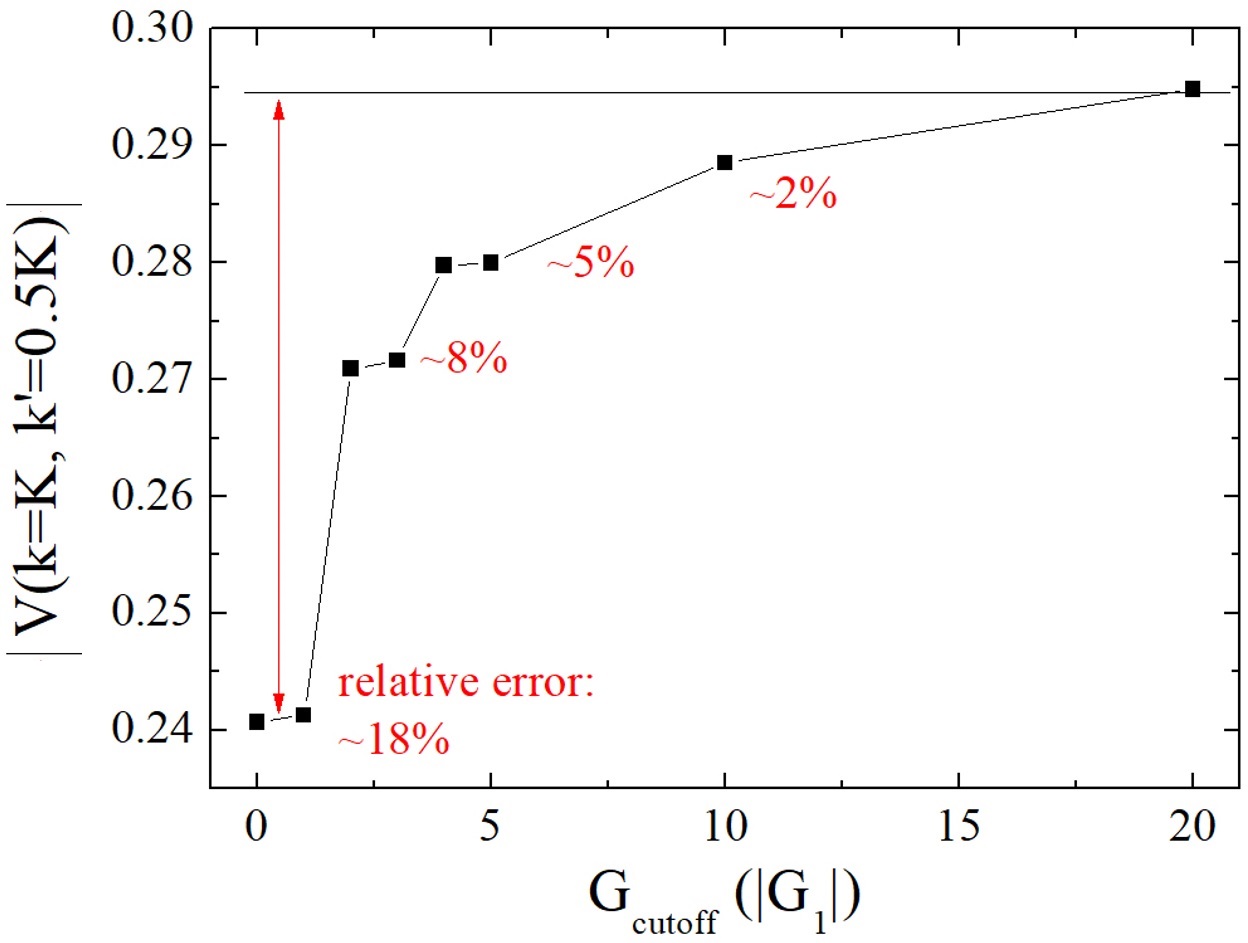}
\caption{Illustration of convergence of absolute value of Coulomb matrix element with respect to summation over the reciprocal lattice vectors $G_{\textrm{cutoff}}$.} 
\label{figA4}
\end{figure}

The numerical evaluation of direct electron-hole  Coulomb matrix elements including full Bloch wavefunctions  with Berry's phases is the most demanding part of our calculations. For example, for a lattice of $\sim 7000$ k-point one needs to estimate around $\sim (7000^2)/2$ matrix elements, giving $\sim 25\cdot 10^6$ elements in total. The calculations of form factors in 
Eq. (\ref{eq15}) entering summations over reciprocal lattice vectors $\vec{G}$ (Eq. (\ref{eq14})) are therefore subject to several numerical convergence parameters. Also, summation over $G$ vectors can be done up to  a finite $G_{\textrm{cutoff}}$, as shown in Fig. (\ref{figA4}). The integration over $z,z'$ in Eq. (\ref{eq15}) is usually performed from $z_{\textrm{min}}=-5.0$ $a_B$ to $z_{\textrm{min}}=-5.0 $ $a_{B}$, where $a_{B}$ is Bohr radius. This approximation is reasonable due to the finite spread of Slater-like localized orbitals. The grid density is usually set to 0.5 $a_B$, showing good convergence of $z,z'$ integrals (compare with Fig. \ref{figA3}). When calculating co-densities,  Eq. (\ref{eq16}), we first check if given pair of tight-binding coefficients $v_{\alpha\mu}^{*}v_{\beta\nu}$ is larger than a cut-off value, usually 0.1, and then we estimate the value of in-plane 2D integral on some coarse grid, improving to higher density grids when significant values of  integrals are found. The number of unit cells in summation in Eq. (\ref{eq16}) is usually 7 (central unit cell + 6 nearest neighbors, each of them containing one Mo and two S atoms). All 2D integration domains are optimized to take into account only orbitals at relevant , for a given integral,  positions $U_i + \tau_\alpha$ and $U_j + \tau_\beta$, with some integration domain off-set, usually $\sim 2.5$$a_B$, again due to finite spread of Slater-like orbitals.

\section*{Appendix F: Coulomb interaction form factors and reduction to simplified interactions form}

\begin{figure}
\centering
\includegraphics[width=0.5\textwidth]{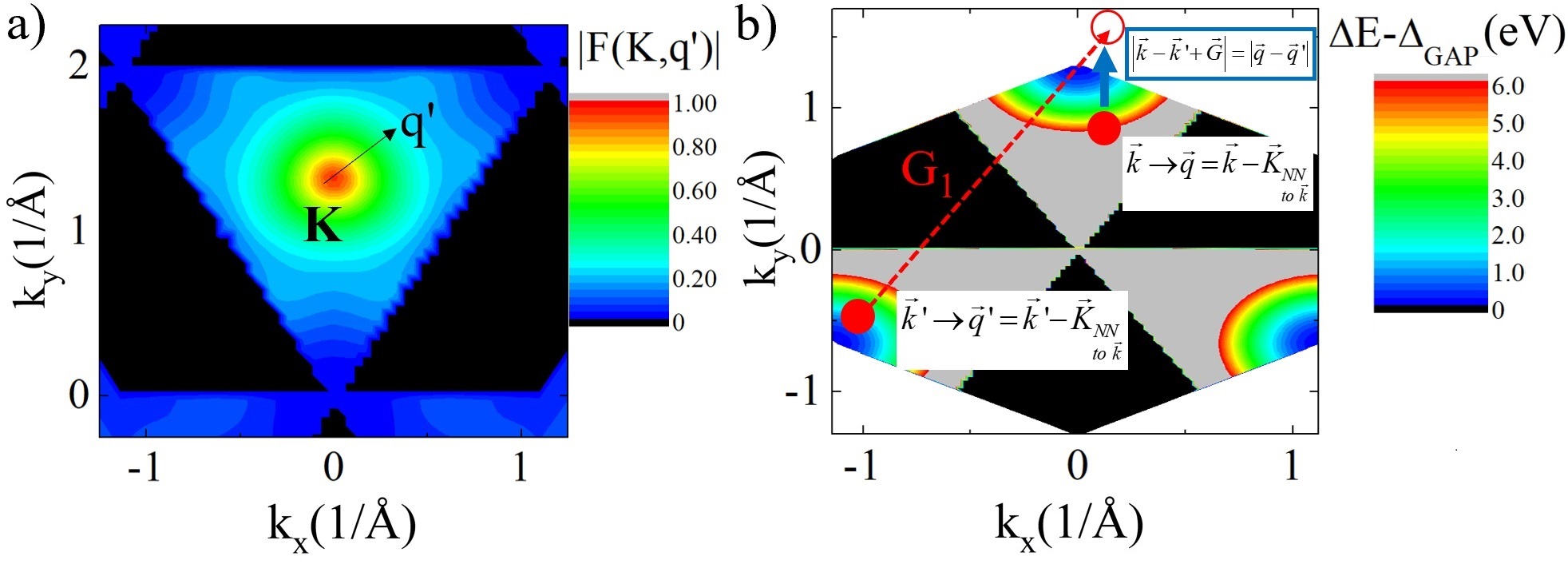}
\caption{(a) Schematic absolute value of direct Coulomb interaction $V^{D}$ form factor $\left|F^{D}\left(\vec{K},\vec{q'},0\right)\right|$ around K-point for $\vec{G}=0$ showing additional $1/|q-q'|$ type contribution to electron-hole interaction from wavefunction effects. (b) Construction of $1/|q-q'|$ interaction on Brillouin zone with respect to neighborhood of K-point, taking into account reciprocal lattice vector translations. } 
\label{figA5}
\end{figure}

The direct electron-hole Coulomb interaction form factors F, Eq. (\ref{eq14}), are important quantities for correct description of the effect of  topology on the excitonic spectrum. In addition, they also describe the three dimensional character of the charge density which contributes strongly to the magnitude of matrix element, modifying the $1/|k-k'-G|$ 2D dependence to  $1/|k-k'-G|^{\eta}$ (see Eq. (\ref{eq14})) instead, resulting in $\eta>1$ correction to overall magnitude of the matrix element, as shown in Fig. (\ref{figA5})(a). This correction is similar to the effect of Rytova-Keldysh screening, therefore when extracting values of polarizability $\alpha$ entering this model from comparison with experimental excitonic series, one has to be careful not to overestimate the effects of dynamical screening in comparison with orbital wavefunctions contributions.

Interestingly, the form factor functions $F(k,k',G)$ maximize their value for such reciprocal lattice vectors $\vec{G}$, that $|k-k'-G|$ distance entering Eq. (\ref{eq14}) is reduced to the distance $|q-q'|$ for q vectors defined by $\vec{k}=\vec{K}+\vec{q}$, where q are vectors measured from nearest $K$ points. Such translation of every pair of (k,k') points to shortest $|q-q'|$ vector allows us to justify the simplified picture of electron-hole interaction, reduced to the neighborhood of one of the K-points, e.g. center of triangular $+K$ valley, as shown in Fig. (\ref{figA5})(b). To be more specific, e.g. for matrix element $V(k=K,k')$ shown in Fig. (\ref{figA5})(a), the largest contributions in summation in Eq. (\ref{eq14}) over G are coming from:
\begin{equation}
\begin{split}
&V(K,k')=\gamma\Bigg(\frac{F(K,k',G=0)}{|k'-K|}+ \\
&\frac{F(K,k',G=-G_{1})}{|k'-K+G_{1}|}+\frac{F(K,k',G=-G_{1}+G_{2})}{|k'-K+G_{1}-G_{2}|} \Bigg),
\end{split}
\end{equation}
for $\vec{G}_{1,2}=\vec{b}_{1,2}$ as defined on Fig. (\ref{fig1})(b). This allows us to write this matrix element around K-point as
\begin{equation}
V(K,q')=\gamma\frac{F(K,q')}{|K-q'|}.
\end{equation}
\\[0.5in]
\bibliography{X-MoS2-28Feb2020}

\end{document}